\documentclass[twocolumns,9pt]{IEEEtran}
\usepackage{cite}
\usepackage{amsmath,amssymb,amsfonts}
\usepackage{algorithmic}
\usepackage{graphicx}
\usepackage{textcomp}

\usepackage{amsmath,epsfig}

\usepackage{amsmath,epsfig,amsfonts,amssymb,graphics,theorem,calc,url,bm}
\usepackage{array}
\usepackage{calc}
\usepackage{verbatim}

\usepackage{amsmath}
\usepackage{amsfonts}
\usepackage{mathrsfs}
\usepackage{pifont}
\usepackage{amssymb}
\usepackage{verbatim}
\usepackage{upgreek}
\usepackage{color}
\usepackage{graphicx}
\usepackage{algorithmic}
\usepackage{algorithm}
\usepackage{epsfig}

\usepackage{fancybox}

\usepackage{amsfonts}
\usepackage{mathrsfs}
\usepackage{pifont}

\usepackage{verbatim}
\usepackage{upgreek}
\usepackage{color}

\usepackage{algorithmic}
\usepackage{algorithm}

\usepackage{fancybox}

\newcommand{\bzero}{\mbox{\boldmath{$0$}}}

\newcommand{\bA}{\mbox{\boldmath{$A$}}}
\newcommand{\ba}{\mbox{\boldmath{$a$}}}
\newcommand{\bB}{\mbox{\boldmath{$B$}}}

\newcommand{\bC}{\mbox{\boldmath{$C$}}}

\newcommand{\bd}{\mbox{\boldmath{$d$}}}

\newcommand{\bI}{\mbox{\boldmath{$I$}}}
\newcommand{\ii}{\mbox{\boldmath $i$}}%%%%%%%%%%%%%%%%%%

\newcommand{\bn}{\mbox{\boldmath{$n$}}}

\newcommand{\bQ}{\mbox{\boldmath{$Q$}}}

\newcommand{\bR}{\mbox{\boldmath{$R$}}}

\newcommand{\bs}{\mbox{\boldmath{$s$}}}

\newcommand{\bW}{\mbox{\boldmath{$W$}}}
\newcommand{\bw}{\mbox{\boldmath{$w$}}}
\newcommand{\bX}{\mbox{\boldmath{$X$}}}
\newcommand{\bx}{\mbox{\boldmath{$x$}}}

\newcommand{\by}{\mbox{\boldmath{$y$}}}
\newcommand{\bZ}{\mbox{\boldmath{$Z$}}}
\newcommand{\bz}{\mbox{\boldmath{$z$}}}

\newcommand{\bDelta}{\mbox{\boldmath{$\Delta$}}}

\newcommand{\tr}{\mbox{\rm tr}\, }

\newcommand{\rank}{\mbox{\rm Rank}\, }

\newcommand{\ex}{{\bf\sf E}}

\def\proof {{\bf Proof:} }
\def\endproof{\hfill $\Box$}% \vskip .5cm}

\newtheorem{theorem}{Theorem}[section]

\newtheorem{lemma}[theorem]{Lemma}
\newtheorem{proposition}[theorem]{Proposition}
\begin{document}
%\markboth{Submitted to IEEE Trans. on Signal Processing}{...}

%\title{Probabilistic Methods for the Global Optimal Solutions of QCQP}
\title{Robust Adaptive Beamforming via Worst-Case SINR Maximization with Nonconvex Uncertainty Sets}
\vspace{2cm}

\author{
Yongwei Huang, {\it Senior Member, IEEE}\thanks{Y. Huang is with School of Information Engineering, Guangdong University of Technology, University Town, Guangzhou, Guangdong 510006, China (email: ywhuang@gdut.edu.cn).},
Hao Fu\thanks{H. Fu is with Huawei Technologies, Software Avenue, Yuhuatai District, Nanjing, Jiangsu, China (email: fuhao33@huawei.com; 2111703038@mail2.gdut.edu.cn).},
Sergiy A. Vorobyov, {\it Fellow, IEEE}\thanks{S. A. Vorobyov is with the Department of Signal Processing and Acoustics, Aalto University, 02150 Espoo, Finland (e-mail: svor@ieee.org).},
Zhi-Quan Luo, {\it Fellow, IEEE}\thanks{Z.-Q. Luo is with School of Science and Engineering, The Chinese University of Hong Kong, Shenzhen, and Shenzhen Research Institute of Big Data, Guangdong 518172, China (email: luozq@cuhk.edu.cn).}
}

\maketitle
\begin{abstract}
This paper considers a formulation of the robust adaptive beamforming (RAB) problem based on worst-case signal-to-interference-plus-noise ratio (SINR) maximization with a nonconvex uncertainty set for the steering vectors. The uncertainty set consists of a similarity constraint and a (nonconvex) double-sided ball constraint. The worst-case SINR maximization problem is turned into a quadratic matrix inequality (QMI) problem using the strong duality of semidefinite programming. Then a linear matrix inequality (LMI) relaxation for the QMI problem is proposed, with an additional valid linear constraint. Necessary and sufficient conditions for the tightened LMI relaxation problem to have a rank-one solution are established. When the tightened LMI relaxation problem still has a high-rank solution, the LMI relaxation problem is further restricted to become a bilinear matrix inequality (BLMI) problem. We then propose an iterative algorithm to solve the BLMI problem that finds an optimal/suboptimal solution for the original RAB problem by solving the BLMI formulations. To validate our results, simulation examples are presented to demonstrate the improved array output SINR of the proposed robust beamformer.
\end{abstract}

\vspace{0.5cm}

%\begin{center}
%\begin{small}
%{\bf Keywords.}
%\end{small}
%\end{center}


%\newpage

\section{Introduction}
Adaptive beamforming has been widely employed in many applications to radar, sonar, communications, microphone array speech/audio processing, medical imaging, radio astronomy, and others \cite{RABapplications-book}. Since traditional adaptive beamforming approaches have weak immunity against even a small mismatch between the presumed and actual signal steering vectors, many robust adaptive beamforming (RAB) techniques have been developed and substantial progress has been made in this area in recent years (see e.g. \cite{Stoica-Li-book-beamforming, Voro13survey, GSSBO09, Gu-tsp12, LHuang-tsp15, PZhang-taes-21} and references therein). The mismatch may result from look direction errors, imperfect array calibration, distorted antenna shape, and other reasons \cite{Stoica-Li-book-beamforming}. Facilitated by the recent development of convex optimization and robust optimization \cite{Boydbook, RO-book-Nemirovski}, new powerful RAB methods have been proposed with significantly improved performance.

Among all proposed RAB techniques, the approach based on the concept of worst-case performance optimization has drawn much attention \cite{Stoica-Li-book-beamforming, Voro13survey, GSSBO09, VGL03tsp, VGLM04spl, Lorenz-Boyd05,Kim08}. Particularly in \cite{VGL03tsp}, the actual signal steering vector is modelled as a sum of the presumed steering vector and a norm bounded mismatch vector, and the worst-case signal-to-interference-plus-noise ratio (SINR) maximization problem is formulated. It turns out that this problem can be reformulated as minimization of the denominator of the SINR subject to infinitely many nonconvex quadratic constraints over a spherical uncertainty set of the mismatch vectors. The (robust) minimization problem can be further transformed equivalently into a second-order cone programming (SOCP) problem, and consequently solved via an interior-point method (see e.g. \cite{IPM-book-Wright}). Both uncertainties in the signal steering vector and in the array signal matrix are considered in \cite{VGLM04spl}, and RAB technique based on the worst-case performance optimization is obtained via an SOCP problem, which is a reformulation for the worst-case SINR maximization over the two (convex) uncertainty sets. In \cite{Kim08}, under the assumption that the steering vector and covariance matrix are mismatched, but known to belong to a general convex compact subset, the problem of finding a beamforming vector that maximizes the worst-case SINR over the uncertainty set is studied. It has been shown that when the uncertainty set can be represented in terms of linear matrix inequalities (LMIs), the worst-case SINR maximization problem can be solved via semidefinite programming (SDP).

In many practical scenarios, it is inadequate to describe the uncertainty sets of steering vectors by only convex constraints. Indeed, some important constraints are nonconvex by nature (e.g., the unit norm constraint). Therefore, it is necessary to study the worst-case SINR maximization problem with nonconvex uncertainty sets, which have practical significance. It can also be viewed as a desired and technically advanced and nontrivial generalization of the work in \cite{Kim08}.

In this paper, we propose a RAB technique that maximizes the worst-case SINR over a nonconvex uncertainty set of the signal-of-interest (SOI) steering vector, with a similarity constraint  and a (nonconvex) double-sided ball constraints. The similarity constraint is a spherical constraint, ensuring the desired steering vectors to be sufficiently close to a presumed steering vector, while the ball constraints are imposed to account for the steering vector gain perturbations caused, for example, by the sensor amplitude and phase errors as well as the sensor position errors (cf. \cite[pp. 2408 and 2414]{jianli04}). Since the uncertainty set is nonconvex, the solution methods proposed herein to the new worst-case SINR maximization problem are very different from that in \cite{Kim08}. In particular, the contributions of this paper are the following.

\begin{itemize}
\item We reformulate the worst-case SINR maximization RAB design problem over the aforementioned nonconvex uncertainty set as a semi-infinite program, namely, a quadratic minimization problem with infinitely many constraints over the nonconvex uncertainty set. Mathematically, the constraints for the uncertainty set consist of an inhomogeneous quadratic constraint and a double-sided quadratic constraint. In this case, a quadratic optimization over such uncertainty set is hidden convex, and its SDP relaxation is tight \cite{HP14mar}. Then the constraint set of the semi-infinite program can be represented by a single quadratic matrix inequality (QMI) constraint by applying the strong duality theory of SDP. The quadratic minimization problem with such QMI constraint is called a QMI problem. Thus, our worst-case SINR maximization problem is also shown to be a QMI problem.

\item For the QMI problem, we establish an LMI relaxation problem. Often there is a gap between an LMI relaxation and a QMI problem. In order to narrow or close the gap, in our case, a valid linear constraint is added to the LMI problem, which means that the relaxation problem is further tightened. Some necessary and sufficient conditions for the tightened LMI problem to have a rank-one solution is presented. If the tightened LMI relaxation problem admits a rank-one solution (output by an interior-point method), then the solution is globally optimal for the QMI problem, and thus, for the original worst-case SINR maximization problem.

\item When the tightened LMI relaxation problem however has a high-rank solution, we impose to the problem an additional rank-one condition, which is characterized by the constraint requiring second largest eigenvalue of a positive semidefinite (PSD) beamforming matrix to be nonpositive. This is a novel approach, and such characterization is universal, thus, such constraint can be added to any problem where one seeks a nonzero rank-one solution. In this way, the tightened LMI relaxation problem can be further restricted. Moreover, the tightened LMI relaxation problem after the restriction is shown to be equivalent to a bilinear matrix inequality (BLMI) problem, which follows from an SDP representation for the sum of $K$ largest eigenvalues of a Hermitian matrix. Such BLMI problem is nonconvex in general. Nevertheless, we propose an iterative algorithm to solve it, which finds an optimal/suboptimal solution for the original QMI problem associated with the worst-case SINR maximization problem.

\item In addition to the aforementioned uncertainty set, we consider another nonconvex uncertainty set of practical significance, which is comprised of a double-sided ball constraint and a quadratic constraint that prevents the SOI's direction-of-arrival (DOA) from converging to the set of DOAs characterized by all linear combinations of the interference steering vectors (cf. \cite{KVH2012tsp, HZV2019tsp}). Then, the worst-case SINR maximization problem over the new uncertainty set is formulated in order to compute a RAB weight vector, and an approach similar to the previous one is developed to find an optimal/suboptimal solution of the problem. This approach includes the steps of QMI problem reformulation, solving a tightened LMI relaxation problem, BLMI problem reformulation, and developing an iterative algorithm for solving the BLMI problem.
\end{itemize}

The remainder of this paper is organized as follows. Section~II is devoted to signal models and problem formulation. In Section~III, we consider the scenario with the uncertainty set consisting of a similarity constraint and a double-sided ball constraint. In Section~IV, we then handle the case with the uncertainty set comprised of a quadratic constraint and the double-sided ball constraint. Numerical examples are presented in Section~V. Finally, conclusions are drawn in Section~VI.

{\it Notation}: We adopt the notation of using boldface for vectors
$\ba$ (lower case), and matrices $\bA$ (upper case). The transpose
operator and the conjugate transpose operator are denoted by the
superscripts $(\cdot)^T$ and $(\cdot)^H$, respectively. The notation $\mbox{tr}(\cdot)$ stands for the trace of a square matrix argument; $\bI$ and ${\bf 0}$
denote respectively the identity matrix and the matrix (or the row
vector or the column vector) with zero entries (their size is
determined from the context). The letter $j$ represents the
imaginary unit (i.e., $j=\sqrt{-1}$), while the letter $i$ often
serves as an index. For any complex number $x$, we use
$\Re(x)$ and $\Im(x)$ to denote respectively the real and
imaginary parts of $x$, $|x|$ and $\mbox{arg}(x)$ represent the
modulus and the argument of $x$, and $x^*$ ($\bx^*$ or $\bX^*$)
stands for the component-wise conjugate of $x$ ($\bx$ or $\bX$).
 %$\lambda_{\max}(\bA)$
%stands for the largest eigenvalue of $\bA$.
%The symbols $\odot$ and $\otimes$ represent
%the Hadamard element-wise and the Kronecker product,
% respectively \cite{Horn85}.
The Euclidean norm of vector $\bx$ is denoted by $\|\bx\|$, and the Frobenius norm (the spectral norm) of
matrix $\bX$ by $\|\bX\|_F$ ($\|\bX\|_2$).
%The symbol $\odot$ represents the Hadamard element-wise product.
The curled inequality symbol $\succeq$ (and its strict form $\succ$)
is used to denote generalized inequality: $\bA\succeq\bB$ meaning that
$\bA-\bB$ is a Hermitian positive semidefinite matrix
($\bA\succ\bB$ for positive definiteness).  The space of Hermitian $N\times N$ matrices
(the space of real-valued symmetric $N\times N$ matrices) is denoted by ${\cal
H}^N$ (${\cal S}^N$), and  the set of all positive semidefinite
matrices in ${\cal H}^N$ (${\cal S}^N$) by ${\cal
H}_+^N$ (${\cal S}_+^N$). The notation $\ex[\cdot]$ represents the
statistical expectation. %, $\vect(\bX)$ denotes the vector stacked by the columns of $\bX$, and $\diag(\bx)$ denotes the diagonal matrix with the diagonal elements being the components of $\bx$.
%The notations $\range(\bX)$ and $\nul(\bX)$ stand for the range space and the null space, respectively.
The notation $\rank(\bX)$ stands for the rank of a matrix.
All eigenvalues $\lambda_n(\bX)$, $n= 1, \cdots, N$ of Hermitian matrix $\bX$ are placed in a descending order, namely, $\lambda_1(\bX)\ge\lambda_2(\bX)\ge\cdots\ge\lambda_N(\bX)$.
Finally, $v^\star(\cdot)$  represents the optimal value of problem $(\cdot)$.
%$\bX^{1/2}$ stands for the square root matrix if $\bX\succeq\bzero$.
%We denote by $\mathbb{R}_+^L$ the set of $L$-dimension nonnegative vectors.

\section{Signal models and problem formulation}\label{Motivation-Applications}
%\subsection{Robust Receive  Beamforming}
Let the narrowband signal received by an $N$-sensor array be given by
\[
\by(t)=\bs(t)+\ii(t)+\bn(t),
\]
where $\bs(t)$, $\ii(t)$, and $\bn(t)$ are statistically independent vectors corresponding to the SOI, interference, and noise, respectively. In the point
signal source case,  $\bs(t)$ is expressed as
\[
\bs(t)=s(t)\ba
\]
where $s(t)$ is
the SOI waveform and $\ba$ is its steering vector (the actual array response or spatial signature of the SOI). The receive beamformer outputs the signal
\[
x(t)=\bw^H\by(t)
\]
where $\bw$ is the $N\times 1$ vector of beamformer complex weight coefficients (called a beamvector). The beamforming problem is to find an optimal  beamvector $\bw$ maximizing the beamformer output SINR
\begin{equation}\label{SINR-def}
\mbox{SINR}=\frac{\sigma_s^2|\bw^H\ba|^2}{\bw^H\bR_{i+n}\bw}
\end{equation}
where $\sigma_s^2$ is the SOI power and $\bR_{i+n}$ is the interference-plus-noise covariance matrix. In practical applications, the true covariance matrix $\bR_{i+n}$ is unavailable and, consequently, the data sample estimate
\begin{equation}\label{sampling-R}
\bar\bR=\frac{1}{T}\sum_{t=1}^T \by(t)\by(t)
\end{equation}
is used instead, with $T$ being the number of available snapshots and $\by(t)$ being the beamformer training data. The SINR maximization problem is equivalent to the following convex problem
\begin{equation}\label{opt-beamforming-vector-nonrobust}
\underset{\bw}{\sf{minimize}}~\bw^H\bar\bR\bw~~ {\sf{subject\;to}}~ |\bw^H\ba|=1,
\end{equation}
and its solution is
\begin{equation}\label{MV-beamformer}
%\bw=e^{j\theta}(\ba_0^H\hat\bR^{-1}\ba_0)^{-1}\hat\bR^{-1}\ba_0, \, \forall \theta\in\mathbb{R}.
\bw^\star=\frac{\bar\bR^{-1}\ba}{\ba^H\bar\bR^{-1}\ba}.
\end{equation}
This is the well-known minimum variance distortionless response (MVDR) beamformer (see e.g. \cite{Lorenz-Boyd05}) or Capon beamformer (see e.g. \cite{jianli04}).%, and the corresponding beamformer output power $\ex[|\bw^{\star H}\by(t)|^2]$ is written as
%\begin{equation}\label{MV-beamformer-output-power}
%\bw^{\star H}\hat\bR\bw^\star=\frac{1}{\ba^H\hat\bR^{-1}\ba}.
%\end{equation}

In practical scenarios (e.g., multi-antenna wireless communications and passive source localization), the beamformer suffers from dramatic performance degradation due to %inaccurate estimate of $\hat\bR$ and
mismatch between the actual steering vector $\ba$ and the presumed steering vector $\hat\ba$. To mitigate the degradation, many robust adaptive beamforming techniques have been proposed in the past two decades  (see e.g. \cite{Stoica-Li-book-beamforming}, \cite{VGL03tsp}, \cite{Lorenz-Boyd05},   \cite{Kim08}, \cite{jianli04}, \cite{KVH2012tsp}, \cite{HZV2019tsp} and references therein).

A popular robust beamforming design is the so-called MVDR robust adaptive beamforming (RAB), based on the optimal estimate of the actual SOI's steering vector $\ba$ in accordance to some criterion (for instance, maximizing the beamformer output power subject to some constraints on $\ba$, see e.g. \cite{KVH2012tsp}, \cite{HZV2019tsp}, \cite{H-V-W08SPL} and references therein). Note that the beamformer output power can be expressed as
\begin{eqnarray}\nonumber
\ex[|x(t)|^2]&=&\ex[\bw^H\by(t)\by(t)^H\bw]\\ \nonumber
                   &=&\bw^H\ex[\by(t)\by(t)^H]\bw\\ \nonumber
                   &\approx&\bw^H\bar\bR\bw.
\end{eqnarray}
Using the MVDR beamformer formula \eqref{MV-beamformer}, the array output power at $\bw^\star$ can be found to be
\[
\bw^{\star H}\bar\bR\bw^\star=\frac{1}{\ba^H\bar\bR^{-1}\ba}.
\]
 In order to maximize the array output power, we just can  minimize the denominator subject to a set of constraints on the actual steering vector. In other words, the following optimal steering vector estimation problem needs to be solved
\begin{equation}\label{opt-est-steering-vector-general}
\underset{\ba\in{\cal A}}{\sf{minimize}}~\ba^H\bar\bR^{-1}\ba,
\end{equation}
where ${\cal A}$ is a constraint set for the actual steering vector.
Once an optimal solution $\ba^\star$ is obtained, we get an optimal MVDR RAB vector $\bw^\star$ by substituting $\ba^\star$ into \eqref{MV-beamformer}. The uncertainty set ${\cal A}$ here can be a nonconvex, for example, (cf. \cite{HZV2019tsp})
\begin{equation}\label{example-A-2}
{\cal A}_1=\{\ba~|~\|\ba-\hat\ba\|^2\le\epsilon,\,N-\eta_1\le\|\ba\|^2\le N+\eta_2\}.
\end{equation}
Note that the MVDR RAB vector $\bw^\star$ (cf. \cite{HZV2019tsp}) can be also interpreted as a solution for the minimax problem: % (as long as $\bar\bR\succ\bzero$):
\begin{equation}\label{min-max-MVDR}
\begin{array}[c]{clcl}
\underset{\ba\in{\cal A}}{\sf{minimize}} & \underset{\bw\ne\bzero}{\sf{maximize}}
\begin{array}[c]{c}\displaystyle\frac{|\bw^H\ba|^2}{\bw^H\bar\bR\bw}\end{array}&=&\underset{\ba\in{\cal A}}{\sf{minimize}}~~\ba^H\bar\bR^{-1}\ba.
\end{array}
\end{equation}
%In \eqref{example-A-2}, it is highlighted that the double-sided norm constraint accounts for  gain perturbations,  sensor position errors, mutual coupling, and phase errors (cf. \cite[pp. 2408 and 2414]{jianli04}).

In the literature, another interesting robust adaptive beamformer is based on the worst-case SINR maximization design. Specifically, the maximization problem of the worst-case SINR is formulated as
\begin{equation}\label{max-min-2003-0}
\begin{array}[c]{cl}
\underset{\bw\ne\bzero}{\sf{maximize}} &\underset{\ba\in{\cal A},\bDelta\in{\cal B}}{\sf{minimize}}
\begin{array}[c]{c}\displaystyle\frac{|\bw^H\ba|^2}{\bw^H(\bar\bR+\bDelta)\bw},
\end{array}
\end{array}%\right.
\end{equation}where ${\cal B}$ is the uncertainty set for matrix error term $\bDelta$.
%(see e.g. \cite{Kim08}, where the uncertainty set is a general convex set).
The robust beamforming problem can be equivalently transformed into
\begin{equation}\label{max-min-2003-1}
\begin{array}[c]{cl}
\underset{\bw\ne\bzero}{\sf{maximize}} &
\begin{array}[c]{c}\displaystyle\frac{\underset{\ba\in{\cal A}}{\sf{minimize}}~|\bw^H\ba|^2}{\underset{\bDelta\in{\cal B}}{\sf{maximize}}~\bw^H(\bar\bR+\bDelta)\bw},
\end{array}
\end{array}%\right.
\end{equation}
which can be further reexpressed as
\begin{equation}\label{max-min-2003-2}
\underset{\bw\ne\bzero}{\sf{min}}~\underset{\bDelta\in{\cal B}}{\sf{max}}~\bw^H(\hat\bR+\bDelta)\bw~~ {\sf{subject\;to}}~ \underset{\ba\in{\cal A}}{\sf{min}}~|\bw^H\ba|^2=1.
\end{equation}
In other words, if a vector $\bw^\star$ is optimal for \eqref{max-min-2003-2}, then it is also optimal for \eqref{max-min-2003-1}, and thus, for \eqref{max-min-2003-0}.
Therefore, the worst-case performance optimization design takes the following formulation
\begin{equation}\label{rob-beamf-0}
\underset{\bw\ne\bzero}{\sf{min}}~\underset{\bDelta\in{\cal B}}{\sf{max}}~\bw^H(\bar\bR+\bDelta)\bw~~ {\sf{subject\;to}}~ |\bw^H\ba|^2\ge1,\,\forall \ba\in{\cal A},
\end{equation}
since both problems \eqref{rob-beamf-0} and \eqref{max-min-2003-2} are tantamount to each other.

In the %seminal
paper \cite{VGL03tsp}, the following  robust adaptive beamforming problem has been studied
\begin{equation}\label{robust-opt-Vor03}
\underset{\bw}{\sf{minimize}}~\bw^H\bar\bR\bw~~ {\sf{subject\;to}}~ |\bw^H\ba|^2\ge1,\,\forall\ba: \|\ba-\hat\ba\|^2\le\epsilon.
\end{equation}
%In words, the set of the actual steering vector is the ball centered at the presumed $\hat\ba$ (with $\|\hat\ba\|^2=N$):
%\begin{equation}\label{uncertainty-set-luo03}
%{\cal A}_0=\{\ba~|~\|\ba-\hat\ba\|\le\epsilon\}.
%\end{equation}
In words, the set ${\cal A}=\{\ba~|~\|\ba-\hat\ba\|\le\sqrt{\epsilon}\}$ is convex and it is a ball centered at a presumed steering vector $\hat\ba$ (with $\|\hat\ba\|^2=N$).
It turned out that the robust optimization problem \eqref{robust-opt-Vor03} can be equivalently transformed into an SOCP of the form
\begin{equation}\label{robust-opt-Vor03-socp-equiv}
\underset{\bw}{\sf{minimize}}~\bw^H\hat\bR\bw~~ {\sf{subject\;to}}~ \Re(\bw^H\bar\ba)\ge\sqrt{\epsilon}\|\bw\|+1.
\end{equation}
%by employing the fact
%\[
%\underset{\|\bdelta\|^2\le\epsilon}{\sf{minimize}}~|\bw^H(\hat\ba+\bdelta)|~=~\max\{|\bw^H\hat\ba|-\sqrt{\epsilon}\|\bw\|,0\}
%\]
%(see \cite{luo03} or \cite[Lemma 3.1]{{HuangLiMaZhang-TSP12}}), and the fact that the objective function
%in \eqref{robust-opt-Luo03} remains the same for  arbitrary phase rotation of $\bw$.

In this paper, we focus on the  robust adaptive beamforming problem \eqref{rob-beamf-0} with  a nonconvex set of possible steering vectors and a convex set of perturbations of the interference-plus-noise covariance matrix, which generalizes the works in \cite{VGL03tsp,Kim08}. %, where convex uncertainty sets are considered.
Specifically, the uncertainty set ${\cal B}$ is defined as
\begin{equation}\label{uncertainty-set-B1}
{\cal B}_1=\{\bDelta~|~\|\bDelta\|_F^2\le\gamma,\,\bar\bR+\bDelta\succeq\bzero\}.
\end{equation}
Therefore, problem \eqref{rob-beamf-0} is rewritten as
\begin{equation}\label{rob-beamf-full}
\underset{\bw\ne\bzero}{\sf{min}}~\bw^H(\bar\bR+\sqrt{\gamma}\bI)\bw~~ {\sf{subject\;to}}~ |\bw^H\ba|^2\ge1,\,\forall \ba\in{\cal A},
\end{equation}since
\begin{equation}\label{bDelta-inequality}
\bw^H\bDelta\bw=\tr(\bDelta\bw\bw^H)\le\|\bDelta\|_F\|\bw\bw^H\|_F\le\sqrt{\gamma}\|\bw\|^2,
\end{equation}and the equalities hold as long as $\bDelta=\sqrt{\gamma}\bw\bw^H/\|\bw\|^2\in{\cal B}_1$. For notational convenience, we let
\begin{equation}\label{INC-extended}
\hat\bR:=\bar\bR+\sqrt{\gamma}\bI,
\end{equation}then the robust adaptive beamforming problem \eqref{rob-beamf-full} can be recast into
\begin{equation}\label{rob-beamf}
\underset{\bw\ne\bzero}{\sf{minimize}}~\bw^H\hat\bR\bw~~ {\sf{subject\;to}}~ |\bw^H\ba|^2\ge1,\,\forall \ba\in{\cal A}.
\end{equation}

As for ${\cal A}$ in \eqref{rob-beamf}, we first assume that it is specified to
\begin{equation}\label{uncertainty-set}
{\cal A}_1=\{\ba~|~\|\ba-\hat\ba\|^2\le\epsilon,\,N-\eta_1\le\|\ba\|^2\le N+\eta_2\},
\end{equation}
namely, \eqref{example-A-2} is considered, where user parameters $\eta_1$ and $\eta_2$ control the norm  bounds of perturbation.  In particular, when $\eta_1=\eta_2=0$, the set of the desired steering vectors reduces to
\begin{equation}\label{uncertainty-set-1-jianli-4}
{\cal A}_1^\prime=\{\ba~|~\|\ba-\hat\ba\|^2\le\epsilon,\,\|\ba\|^2= N\}.
\end{equation}
%which is different from the uncertainty set in \eqref{robust-opt-Vor03}.
%where the constraint set has been considered in \cite{jianli04}.
%In contrast, we will investigate the worst-case SINR maximization problem over nonconvex uncertainty set \eqref{uncertainty-set} of the steering vectors.
Also, we assume subsequently that $\|\hat\ba\|^2=N>\epsilon$ (for $\epsilon\in[N, 2N]$, the discussion is similar).

Second, we assume (cf. \cite{KVH2012tsp,HZV2019tsp}) that
\begin{equation}\label{uncertainty-set-KVH2012}
{\cal A}_2=\{\ba~|~\ba^H\bar\bC\ba\le \Delta_0,\,N-\eta_1\le\|\ba\|^2\le N+\eta_2\},
\end{equation}
and
\begin{equation}\label{uncertainty-set-HZV2019}
{\cal A}_3=\{\ba~|~\ba^H\bC\ba\ge \Delta_1,\,N-\eta_1\le\|\ba\|^2\le N+\eta_2\}.
\end{equation}
Here, the matrix parameter in \eqref{uncertainty-set-KVH2012} $\bar\bC=\int_{\bar\Theta}\bd(\theta)\bd^H(\theta)d\theta$, $\bd(\theta)$ is the steering vector associated with direction $\theta$ that has the structure defined by the antenna array geometry, and $\bar\Theta$ is the complement of an angular sector $\Theta$, where the desired signal is located. Also, parameter $\Delta_0$ is equal to  $\max_{\theta\in\Theta} \bd^H(\theta) \bar\bC \bd(\theta)$. %For more details, we refer to \cite{KVH2012tsp}.
Similarly, matrix parameter $\bC=\int_{\Theta}\bd(\theta)\bd^H(\theta)d\theta$, and the threshold value $\Delta_1=\min_{\theta\in\Theta}\bd^H(\theta)\bC\bd(\theta)$. Both the first constraint in \eqref{uncertainty-set-KVH2012} and that in \eqref{uncertainty-set-HZV2019} are used to avoid the convergence of the steering vector $\ba$ to any  linear combinations of the interference steering vectors (see e.g. \cite[Fig. 2]{HZV2019tsp}). For more motivations and interpretations of the uncertainty sets in \eqref{uncertainty-set-KVH2012} and \eqref{uncertainty-set-HZV2019}, we refer respectively to \cite{KVH2012tsp} and \cite{HZV2019tsp}, where MVDR RAB designs have been studied.
%In \cite{HZV2019tsp}, the authors have demonstrated that the performance of MVDR robust beamforming with the constraints in \eqref{uncertainty-set-HZV2019} often is better than that with the constraints in \eqref{uncertainty-set-KVH2012}.
In this paper, we will consider the maximization of the worst-case SINR over nonconvex uncertainty sets  \eqref{uncertainty-set-KVH2012} and \eqref{uncertainty-set-HZV2019}.

%\begin{eqnarray}\label{radar-code.a}
%\underset{\bc\in\mathbb{C}^N}{\sf{maximize}} \; & &  \bc^H\bR\bc  \\ \label{radar-code.b}
%\sf{subject\;to}               & &   1-\eta_1\le\bc^H\bc\le 1+\eta_2,\\ \label{radar-code.c}
% & & {\bc}^{H}{\bR_1}{\bc}\ge\delta_a,\\ \label{radar-code.d}
% & & \|{\bc}-{\bc_0}\|^2 \le \epsilon,
%\end{eqnarray}
%\end{subequations}

We remark that a trivial upper bound for problem \eqref{max-min-2003-0} is the optimal value for the following minimax problem (due to the weak duality theorem):
\begin{equation}\label{min-max-2003-0}
\begin{array}[c]{cl}
\underset{\ba\in{\cal A},\bDelta\in{\cal B}}{\sf{minimize}} & \underset{\bw\ne\bzero}{\sf{maximize}}
\begin{array}[c]{c}\displaystyle\frac{|\bw^H\ba|^2}{\bw^H(\bar\bR+\bDelta)\bw}.
\end{array}
\end{array}
\end{equation}
It follows from \cite{Kim08} that if both ${\cal A}$ and ${\cal B}$ are compact and convex sets, then the two problems \eqref{max-min-2003-0} and \eqref{min-max-2003-0} are equivalent to each other, and can be solved via convex optimization. However, when ${\cal A}$ and ${\cal B}$ are specified to ${\cal A}_1$ (or ${\cal A}_2$ or ${\cal A}_3$) and ${\cal B}_1$, respectively,  \eqref{max-min-2003-0} and \eqref{min-max-2003-0} are not equivalent any more since the uncertainty sets ${\cal A}_i$, $i=1,2,3$, are nonconvex.
%In other words, $v_2^\star> v_1^\star$, where $v_1^\star$ and $v_2^\star$ are the  optimal values for \eqref{max-min-2003-0} and \eqref{min-max-2003-0}, respectively. In this case, $(\bw_2^\star,v_2^\star)$ (with $\bw^\star$ being the solution component of an optimal solution $(\ba_2^\star,\bDelta_2^\star,\bw_2^\star)$ for \eqref{min-max-2003-0}) is not robust feasible for the following robust optimization problem
%\begin{equation}\label{max-min-2003-0-equiv}
%\begin{array}[c]{cl}
%\underset{\bw,\,t}{\sf{maximize}} &
%\begin{array}[c]{c} t \end{array}\\
%\sf{subject\;to} &\displaystyle \frac{|\bw^H\ba|^2}{\bw^H(\bar\bR+\bDelta)\bw}\ge t,\,\forall\ba\in{\cal A},\,\bDelta\in{\cal B},
%\end{array}
%\end{equation}which is an equivalent reformulation for the worst-case SINR maximization problem \eqref{max-min-2003-0}. Therefore, when the beamforming vector $\bw_2^\star$ is applied to the array to test the array performance, the real array output SINR may not be higher than that for $\bw_1^\star$ (where $(\ba_1^\star,\bDelta_1^\star,\bw_1^\star)$ is optimal for  \eqref{max-min-2003-0} and thus $(\bw_1^\star,v_1^\star)$ is optimal for \eqref{max-min-2003-0-equiv}), since $\bw_2^\star$ is not sufficiently robust such that the array performance can be degraded.

\section{Solving robust optimization problem \eqref{rob-beamf} with uncertainty set ${\cal A}_1$ in \eqref{uncertainty-set}}
In this section, we show how to convert  robust adaptive beamforming design problem \eqref{rob-beamf} into a QMI problem, and present a method for finding a solution for a tightened LMI relaxation for such QMI problem.

\subsection{QMI Reformulation for Optimization Problem \eqref{rob-beamf} with Uncertainty Set ${\cal A}_1$}
Since the original worst-case SINR maximization problem \eqref{max-min-2003-0} is equivalent to problem \eqref{rob-beamf}, we need only to focus on solving \eqref{rob-beamf} with a specific uncertainty set ${\cal A}$.

To begin with, let us look into the minimization problem for the feasible set in \eqref{rob-beamf} with ${\cal A}={\cal A}_1$, that is,
\begin{equation}\label{min-constraint}%\mbox{(problem name)}%\left\{
\begin{array}[c]{ll}
\underset{\ba\in\mathbb{C}^N}{\sf{minimize}}  & \begin{array}[c]{c}\ba^H\bw\bw^H\ba\end{array}\\
\sf{subject\;to}               &
\begin{array}[t]{l}
\|\ba-\hat\ba\|^2\le\epsilon,\\
N-\eta_1\le\|\ba\|^2\le N+\eta_2.
\end{array}
\end{array}%\right.
\end{equation}
%By the way, we assume that
%\begin{equation}\label{a-hat-esp}
%\|\hat\ba\|^2>\epsilon,
%\end{equation}
%i.e., $N>\epsilon$. The assumption is mild since it just makes the similarity constraint meaningful (otherwise $\|\hat\ba\|^2\le\epsilon$, ).

%{\color{red}By the way, if the problem reduces to
%\begin{equation}\label{min-constraint-reduced}%\mbox{(problem name)}%\left\{
%\begin{array}[c]{ll}
%\underset{\ba\in\mathbb{C}^N}{\sf{minimize}}  & \begin{array}[c]{c}\ba^H\bw\bw^H\ba\end{array}\\
%\sf{subject\;to}               &
%\begin{array}[t]{l}
%\|\ba-\hat\ba\|^2\le\epsilon^2,\\
%%N-\eta_1\le\|\ba\|^2\le N+\eta_2.
%\end{array}
%\end{array}%\right.
%\end{equation}
%then to avoid the trivial case of $\ba^\star=\bzero$ (it is a steering vector), we assume
%\[
%\|\hat\ba\|>\epsilon.
%\]
%In fact, if $\ba^\star=\bzero$, then $\|\hat\ba\|\le \epsilon$. Therefore, if $\|\hat\ba\|> \epsilon$, then $\ba^\star\ne\bzero$.
%}

The optimization problem \eqref{min-constraint} amounts to the following homogeneous quadratically constrained quadratic program (QCQP) (see, e.g., \cite[equation (32)]{HP14mar})
\begin{equation}\label{min-constraint-homo}%\mbox{(problem name)}%\left\{
\begin{array}[c]{cl}
\underset{\bx\in\mathbb{C}^{N+1}}{\sf{minimize}}  & \begin{array}[c]{c}\bx^H\bA_0\bx \end{array}\\
\sf{subject\;to}               &
\begin{array}[t]{l} \bx^H\bA_1\bx \le0,\\
N-\eta_1\le\bx^H\bA_2\bx\le N+\eta_2,\\
\bx^H\bA_3\bx=1,
\end{array}
\end{array}%\right.
\end{equation}
with the matrix parameters being
\[
\bA_0=\left[\begin{array}{cc}\bw\bw^H&\bzero\\ \bzero&0\end{array}\right],\,\bA_1=\left[\begin{array}{cc}\bI&-\hat    \ba\\ -\hat\ba^H& \|\hat\ba\|^2-\epsilon\end{array}\right],
\]
and
\[
\bA_2=\left[\begin{array}{cc}\bI&\bzero\\ \bzero&0\end{array}\right],\,
\bA_3=\left[\begin{array}{cc}\bzero&\bzero\\ \bzero&1\end{array}\right].
\]
The equivalence between \eqref{min-constraint} and \eqref{min-constraint-homo} is in the sense that they share the same optimal value and if $\bx^\star=[\bw^{\star T}, t^\star]^T\in\mathbb{C}^{N+1}$ solves \eqref{min-constraint-homo}, then $\bw^\star/t^\star $ solves \eqref{min-constraint}.

Accordingly, the conventional SDP relaxation of \eqref{min-constraint} is
\begin{equation}\label{min-constraint-homo-SDR}%\mbox{(problem name)}%\left\{
\begin{array}[c]{cl}
\underset{\bX\in{\cal H}^{N+1}}{\sf{minimize}}  & \begin{array}[c]{c}\tr(\bA_0\bX) \end{array}\\
\sf{subject\;to}               &
\begin{array}[t]{l} \tr (\bA_1\bX) \le0,\\
N-\eta_1\le\tr(\bA_2\bX)\le N+\eta_2,\\
\tr(\bA_3\bX)=1,\\
\bX\succeq\bzero.
\end{array}
\end{array}%\right.
\end{equation}
Moreover, the dual problem is the following SDP:
\begin{equation}\label{min-constraint-homo-SDR-dual}%\mbox{(problem name)}%\left\{
\begin{array}[c]{cl}
\underset{\{y_i\}_{i=1}^4}{\sf{maximize}}  & \begin{array}[c]{c}(N-\eta_1)y_2+(N+\eta_2)y_3+y_4 \end{array}\\
\sf{subject\;to}               &
\begin{array}[t]{l} \bA_0-y_1\bA_1-(y_2+y_3)\bA_2-y_4\bA_3\succeq\bzero,\\
y_1\le0,\,y_2\ge0,\,y_3\le0,\,y_4\in\mathbb{R}.
\end{array}
\end{array}%\right.
\end{equation}
%complementary conditions:
%It is known that the complementary conditions (the complement slackness in the first-order optimality condition) are:
%\begin{eqnarray}\label{complmentary-conditions-1}
%0&=&\tr((\bA_0-y_1\bA_1-(y_2+y_3)\bA_2-y_4\bA_3)\bX) \\ \label{complmentary-conditions-2}
%0&=&y_1\tr(\bA_1\bX)\\ \label{complmentary-conditions-3}
%0&=&y_2(\tr(\bA_2\bX)-(N-\eta_1))\\ \label{complmentary-conditions-4}
%0&=&y_3(\tr(\bA_2\bX)-(N+\eta_2))
%\end{eqnarray}
%Note that the scenario of both $y_2>0$ and $y_3<0$ cannot happen due to the nature of the double-sided constraint (otherwise, $\tr(\bA_2\bX)=N-\eta_1=N+\eta_2$, which is a contradiction).

Given the primal and dual SDPs, we claim strict feasibility of both SDPs in the following lemma.
\begin{lemma}\label{strict-feasibility-SDPs-primal-dual}
It holds that the primal SDP \eqref{min-constraint-homo-SDR} and the dual SDP \eqref{min-constraint-homo-SDR-dual} are strictly feasible.
\end{lemma}
\proof
See Appendix~\ref{strict-feasibility-SDPs-primal-dual-proof-appendix}.
\endproof

It follows then from the strong duality theorem (see, e.g., \cite[Theorem 2.4.1]{Nemi-book2001}) that both SDPs are solvable.\footnote{Here by ``solvable", we mean that the minimization (or maximization) problem is feasible, bounded below (above) and the optimal value is attained \cite[page 2]{Nemi-book2001}.} Furthermore, one can always find a rank-one solution for \eqref{min-constraint-homo-SDR} (see e.g. \cite{HP14mar}). In other words, all the optimal values are equal, that is,
\[
v^\star(\eqref{min-constraint})=v^\star(\eqref{min-constraint-homo})=v^\star(\eqref{min-constraint-homo-SDR})=v^\star(\eqref{min-constraint-homo-SDR-dual}).
\]

Accordingly, the robust adaptive beamforming problem \eqref{rob-beamf} with ${\cal A}={\cal A}_1$ is recast into the following QMI problem
%\begin{equation}\label{rob-beamf-QMI}%\mbox{(problem name)}%\left\{
%\begin{array}[c]{ll}
%\underset{\bw,\,\{y_i\}_{i=1}^4}{\sf{minimize}} & \bw^H\hat\bR\bw\\ %\begin{array}[c]{c}\bw^H\hat\bR\bw \end{array}\\
%\sf{s.t.}              & \!\!\!\!\!\!\!\!\!\!\!\!\!\!\! (N-\eta_1)y_2+(N+\eta_2)y_3+y_4=1,\\
%%\bA_0-y_1\bA_1-(y_2+y_3)\bA_2-y_4\bA_3\succeq\bzero,\\
% & \!\!\!\!\!\!\!\!\!\!\!\!\!\!\! \left[\begin{array}{cc}\bw\bw^H-(y_1+y_2+y_3)\bI& y_1\hat\ba\\ y_1\hat\ba^H& -y_4-y_1(\|\hat\ba\|^2-\epsilon )\end{array}\right]\succeq\bzero,\\
% & \!\!\!\!\!\!\!\!\!\!\!\!\!\!\! \bw\in\mathbb{C}^N,\,y_1\le0,\,y_2\ge0,\,y_3\le0,\,y_4\in\mathbb{R}.
%\end{array}%\right.
%\end{equation}
\begin{subequations}\label{rob-beamf-QMI}
\begin{align}\label{rob-beamf-QMI.a}
\underset{\bw,\,\{y_i\}_{i=1}^4}{\sf{minimize}}~~~~ & \bw^H\hat\bR\bw\\ \label{rob-beamf-QMI.b}%\begin{array}[c]{c}\bw^H\hat\bR\bw \end{array}\\
\sf{s.t.} ~~~~~~~~             & \!\!\!\!\!\!\!\!\!\!\!\!\!\!\! (N-\eta_1)y_2+(N+\eta_2)y_3+y_4=1,\\ \label{rob-beamf-QMI.c}
%\bA_0-y_1\bA_1-(y_2+y_3)\bA_2-y_4\bA_3\succeq\bzero,\\
 & \!\!\!\!\!\!\!\!\!\!\!\!\!\!\! \left[\begin{array}{cc}\bw\bw^H-(y_1+y_2+y_3)\bI& y_1\hat\ba\\ y_1\hat\ba^H& -y_4-y_1(\|\hat\ba\|^2-\epsilon )\end{array}\right]\succeq\bzero,\\ \label{rob-beamf-QMI.d}
 & \!\!\!\!\!\!\!\!\!\!\!\!\!\!\! \bw\in\mathbb{C}^N,\,y_1\le0,\,y_2\ge0,\,y_3\le0,\,y_4\in\mathbb{R}.
\end{align}%\right.
\end{subequations}
The conventional LMI relaxation problem can be formulated as
\begin{subequations}\label{rob-beamf-QMI-1-LMI}%\mbox{(problem name)}%\left\{
\begin{align}\label{rob-beamf-QMI-1-LMI.a}
\underset{\bW,\,\{y_i\}_{i=1}^4}{\sf{minimize}}~~~~  &  \tr(\hat\bR\bW) \\ \label{rob-beamf-QMI-1-LMI.b}
\sf{s.t.}~~~~~~~~ & \!\!\!\!\!\!\!\!\!\!\!\!\!\!\! (N-\eta_1)y_2+(N+\eta_2)y_3+y_4=1,\\ \label{rob-beamf-QMI-1-LMI.c}
  & \!\!\!\!\!\!\!\!\!\!\!\!\!\!\! \left[\begin{array}{cc}\bW-(y_1+y_2+y_3)\bI & y_1\hat\ba\\ y_1\hat\ba^H& -y_4-y_1(\|\hat\ba\|^2-\epsilon )\end{array}\right]\succeq\bzero,\\ \label{rob-beamf-QMI-1-LMI.d}
  & \!\!\!\!\!\!\!\!\!\!\!\!\!\!\! \bW\succeq\bzero,\,y_1\le0,\,y_2\ge0,\,y_3\le0,\,y_4\in\mathbb{R}.
\end{align}%\right.
\end{subequations}
%The dual problem of it can be built as follows.
%\begin{equation}\label{rob-beamf-QMI-dual}%\mbox{(problem name)}%\left\{
%\begin{array}[c]{ll}
%\underset{\bZ,\,\bz_1,\,z_0}{\sf{maximize}}  & \begin{array}[c]{c}z_0\end{array}\\
%\sf{subject\;to}
%            & \begin{array}[t]{l}
%\hat\bR-\bZ\succeq\bzero,\\
%\tr\bZ-2\Re(\hat\ba^H\bz_1)+z_0(\|\hat\ba\|^2-\epsilon )\le0,\\
%\tr\bZ-(N-\eta_1)z_0\ge0,\\
%\tr\bZ-(N+\eta_2)z_0\le0,\\
%\left[ \begin{array}{cc} \bZ&\bz_1 \\ \bz_1^H & z_0 \end{array}\right]\succeq\bzero.
%\end{array}
%\end{array}%\right.
%\end{equation}

Suppose that problem \eqref{rob-beamf-QMI-1-LMI} is solvable, %It follows from the strong duality theorem that a sufficient condition for the solvability is: (i) the primal problem  \eqref{rob-beamf-QMI-1-LMI} is feasible and, (ii) the dual problem \eqref{rob-beamf-QMI-dual} is strictly feasible.
 and that $(\bW^\star,\{y_i^\star\}_{i=1}^4)$ is an optimal solution. If $\bW^\star$ is of rank one, then we claim that robust adaptive beamforming problem \eqref{rob-beamf} is solved and the optimal beamforming vector is $\bw^\star$ such that $\bW^\star=\bw^\star\bw^{\star H}$. If rank of $\bW^\star$ is larger than one, then we have to find another rank-one solution for \eqref{rob-beamf-QMI-1-LMI}.%, which represents a nontrivial part in this paper. %We herein study the hard case.

%We note that the variable $y_4$ can be eliminated by substituting the first constraint into the second one.

%It can be further transformed into the problem by introducing an additional variable:
%\begin{equation}\label{rob-beamf-QMI-1}%\mbox{(problem name)}%\left\{
%\begin{array}[c]{ll}
%\underset{\bw,\,\{y_i\}}{\sf{minimize}}  & \begin{array}[c]{c} y_0 \end{array}\\
%\sf{subject\;to}               &
%\begin{array}[t]{l}
%\bw^H\hat\bR\bw\le y_0,\\
%(N-\eta_1)y_2+(N+\eta_2)y_3+y_4=1,\\
%\left[\begin{array}{cc}\bw\bw^H-(y_1+y_2+y_3)\bI& y_1\hat\ba\\ y_1\hat\ba^H& -y_4-y_1(\|\hat\ba\|^2-\epsilon^2)\end{array}\right]\succeq\bzero,\\
%y_1\le0,\,y_2\ge0,\,y_3\le0,\,y_0,y_4\in\mathbb{R}.
%\end{array}
%\end{array}%\right.
%\end{equation}
%In words, the robust adaptive beamforming problem \eqref{rob-beamf} is equivalent to the above QMI problem.

%However, the problem is hard to solve since the the third constraint is a QMI, which in general is non-convex.

\subsection{Tightened LMI Rlelaxation by Adding One More Valid Linear Constraint}
In this subsection, we study how to impose one more constraint to \eqref{rob-beamf-QMI-1-LMI}, such that the feasible region of the LMI relaxation problem  is reduced and it is more likely that the optimal solution is  of rank one. First, observe the lemma below.
\begin{lemma}\label{2nd-large-eigv-nonnegatiive}
The second constraint of \eqref{rob-beamf-QMI} implies that
\begin{equation}\label{additional-linear-constr}
y_1 + y_2 + y_3 \le 0 .
\end{equation}
\end{lemma}
%\proof
%See Appendix \ref{proof-2nd-large-eigv-nonnegatiive}.
%\endproof
The proof is immediate, and thus, is omitted.

In other words, the set of optimal solutions for \eqref{rob-beamf-QMI} will not be altered if one enforces the  linear constraint \eqref{additional-linear-constr} to the problem, that is,
\begin{equation}\label{rob-beamf-QMI-restricted}%\mbox{(problem name)}%\left\{
\begin{array}[c]{ll}
\underset{\bw,\,\{y_i\}_{i=1}^4}{\sf{minimize}}  & \bw^H\hat\bR\bw\\
\sf{subject\;to} & \eqref{rob-beamf-QMI.b},\eqref{rob-beamf-QMI.c},\eqref{rob-beamf-QMI.d},\eqref{additional-linear-constr}\mbox{ satisfied}.\\
%          & y_1+y_2+y_3\le0. \\
\end{array}%\right.
\end{equation}
This problem shares the same optimal value with problem \eqref{rob-beamf-QMI}.

The LMI relaxation problem for \eqref{rob-beamf-QMI-restricted} can be written as
\begin{equation}\label{rob-beamf-QMI-1-LMI-restricted}%\mbox{(problem name)}%\left\{
\begin{array}[c]{ll}
\underset{\bW,\,\{y_i\}_{i=1}^4}{\sf{minimize}}  & \begin{array}[c]{c} \tr(\hat\bR\bW) \end{array}\\
\sf{subject\;to}               &
\begin{array}[t]{l}
%\tr(\hat\bR\bW)\le y_0,\\
\eqref{rob-beamf-QMI-1-LMI.b},\eqref{rob-beamf-QMI-1-LMI.c},\eqref{rob-beamf-QMI-1-LMI.d},\eqref{additional-linear-constr}\mbox{ satisfied}.%\\
% y_1+y_2+y_3\le0.
\end{array}
\end{array}%\right.
\end{equation}
Clearly, the difference between \eqref{rob-beamf-QMI-1-LMI-restricted} and \eqref{rob-beamf-QMI-1-LMI} lies in the additional linear constraint. Therefore, LMI relaxation \eqref{rob-beamf-QMI-1-LMI-restricted} appears tighter than \eqref{rob-beamf-QMI-1-LMI}.
More importantly, if a rank-one solution $\bW^\star=\bw^\star\bw^{\star H}$ is optimal for \eqref{rob-beamf-QMI-1-LMI-restricted}, then $\bw^\star$ is optimal for \eqref{rob-beamf-QMI-restricted}, and thus, also for \eqref{rob-beamf-QMI}. Therefore, let us focus hereafter on tightened LMI relaxation problem \eqref{rob-beamf-QMI-1-LMI-restricted}.  %Interestingly, it is demonstrated in our extensive simulations that the restricted LMI relaxation problem \eqref{rob-beamf-QMI-1-LMI-restricted} always yields a rank-one solution, while the traditional relaxation problem \eqref{rob-beamf-QMI-1-LMI} outputs rank-two solutions.

%\subsection{Sufficient Conditions for rank-one solutions for LMI Relaxation problem \eqref{rob-beamf-QMI-1-LMI-restricted}}

It is not hard to derive the dual problem of \eqref{rob-beamf-QMI-1-LMI-restricted} as
\begin{equation}\label{rob-beamf-QMI-1-LMI-restricted-dual}%\mbox{(problem name)}%\left\{
\begin{array}[c]{ll}
\underset{\bZ,\,\bz_1,\,z_0,\,x}{\sf{maximize}}  & \begin{array}[c]{c}z_0\end{array}\\
\sf{subject\;to}
            & \begin{array}[t]{l}
\hat\bR-\bZ\succeq\bzero,\\
\tr\bZ-2\Re(\hat\ba^H\bz_1)+z_0(\|\hat\ba\|^2-\epsilon )\le x,\\
\tr\bZ-(N-\eta_1)z_0\ge x,\\
\tr\bZ-(N+\eta_2)z_0\le x,\\
\left[ \begin{array}{cc} \bZ&\bz_1 \\ \bz_1^H & z_0 \end{array}\right]\succeq\bzero,\,x\le0.
\end{array}
\end{array}%\right.
\end{equation}

Suppose that the primal and dual SDPs are solvable, and possess the same optimal value. Thus, the complementary conditions include
\begin{subequations}\label{QMI-complementary-conditions}%\mbox{(problem name)}%\left\{
\begin{align}\label{QMI-complementary-conditions-1}
\tr((\hat\bR-\bZ)\bW)&=0,\\ \label{QMI-complementary-conditions-2}
y_1(\tr\bZ-2\Re(\hat\ba^H\bz_1)+z_0(\|\hat\ba\|^2-\epsilon )-x)&=0, \\ \label{QMI-complementary-conditions-3}
y_2(\tr\bZ-(N-\eta_1)z_0- x)&=0, \\ \label{QMI-complementary-conditions-4}
y_3(\tr\bZ-(N+\eta_2)z_0- x)&=0, \\ \label{QMI-complementary-conditions-5}
(y_1+y_2+y_3) x &=0,\\ \label{QMI-complementary-conditions-6}
 \tr\left(\bQ\left[ \begin{array}{cc} \bZ&\bz_1 \\ \bz_1^H & z_0 \end{array}\right] \right)&=0,
\end{align}
\end{subequations}
where
\begin{equation}\label{QMI-complementary-conditions-6-1}
\bQ=\left[\begin{array}{cc}\bW-(y_1+y_2+y_3)\bI& y_1\hat\ba\\ y_1\hat\ba^H& -y_4-y_1(\|\hat\ba\|^2-\epsilon )\end{array}\right].
\end{equation}
It follows from \eqref{QMI-complementary-conditions-1}-\eqref{QMI-complementary-conditions-6} that
\begin{equation}\label{strong-duality-two-LMIs}
\tr(\hat\bR\bW)=z_0,
\end{equation}
i.e., the primal optimal value is equal to the dual optimal value.

Suppose that $(\bW^\star,y_1^\star,y_2^\star,y_3^\star,y_4^\star)$ and $(\bZ^\star,\bz_1^\star,z_0^\star,x^\star)$ are the primal and the dual SDP solutions, respectively. They clearly comply with the complementary conditions \eqref{QMI-complementary-conditions-1}-\eqref{QMI-complementary-conditions-6}. In order to make the notations simpler and clearer, we drop the superscript $^\star$ for the optimal solutions, without leading to confusion. The following propositions are in order.

\begin{proposition}\label{solution-property-1}
Suppose that $(\bW,y_1,y_2,y_3,y_4)$ and $(\bZ,\bz_1,z_0,x)$ are the solutions for primal SDP \eqref{rob-beamf-QMI-1-LMI-restricted} and  dual SDP \eqref{rob-beamf-QMI-1-LMI-restricted-dual}, respectively. Then, it holds that
\begin{enumerate}
\item $y_1<0$;
\item $y_2y_3=0$;
\item if $y_1+y_2+y_3=0$, then $y_1=-y_2$ and $y_3=0$.
\end{enumerate}
\end{proposition}
\proof
See Appendix \ref{proof-solution-property-1}.
\endproof

\begin{proposition} \label{solution-property-2}
Suppose that $(\bW,y_1,y_2,y_3,y_4)$ and $(\bZ,\bz_1,z_0,x)$ are the solutions for primal SDP \eqref{rob-beamf-QMI-1-LMI-restricted} and  dual SDP \eqref{rob-beamf-QMI-1-LMI-restricted-dual}, respectively. Then, it holds that
\begin{enumerate}
\item $z_0=\frac{2\Re(\hat\ba^H\bz_1)-(\tr\bZ-x)}{\|\hat\ba\|^2-\epsilon}$ or $z_0=\frac{y_1x+(y_2+y_3)\tr\bZ}{1-y_4}$ for $y_4\ne1$;
\item $-y_4-y_1(\|\hat\ba\|^2-\epsilon)>0$;
\item if $y_1+y_2+y_3=0$, then $y_2>\frac{1}{2N-\epsilon-\eta_1}$.
\end{enumerate}
\end{proposition}
\proof
See Appendix \ref{proof-solution-property-2}.
\endproof

Note that the second constraint (the LMI constraint) in \eqref{rob-beamf-QMI-1-LMI-restricted} can be recast into
\begin{equation}\label{LMI-constraint-reformulate-Schur-complement}
\bW-(y_1+y_2+y_3)\bI\succeq \bar\ba\bar\ba^H,
\end{equation}
where
\begin{equation}\label{LMI-constraint-reformulate-Schur-complement-a-bar-def}
\bar\ba=\frac{y_1\hat\ba}{\sqrt{-y_4-y_1(\|\hat\ba\|^2-\epsilon)}}.
\end{equation}
%From \eqref{QMI-complementary-conditions-2}, it follows that the second inequality constraint in  dual problem \eqref{rob-beamf-QMI-1-LMI-restricted-dual} is active:
Accordingly, a necessary condition for the existence of rank-one solution for problem \eqref{rob-beamf-QMI-1-LMI-restricted} can be established as follows.

\begin{theorem} \label{solution-property-3-necessary-condition-1}
Suppose that $(\bw\bw^H,y_1,y_2,y_3,y_4)$ and $(\bZ,\bz_1,z_0,x)$ are the solutions for primal SDP \eqref{rob-beamf-QMI-1-LMI-restricted} and dual SDP \eqref{rob-beamf-QMI-1-LMI-restricted-dual}, respectively. Then, it holds that
\begin{enumerate}
\item $(\hat\bR-\bZ)\bw=\bzero$;
\item $y_1+y_2+y_3\le-\frac{\|\bar\ba\|^2-\|\bw\|^2+\sqrt{(\|\bar\ba\|^2+\|\bw\|^2)^2-4|\bar\ba^H\bw|^2}}{2}<0$, if $\bar\ba$ and $\bw$ are linearly independent.
\end{enumerate}
\end{theorem}
\proof
See Appendix \ref{proof-solution-property-3-necessary-condition-1}.
\endproof

In what follows, we present  sufficient conditions for the existence of rank-one solution for problem \eqref{rob-beamf-QMI-1-LMI-restricted}.
 To facilitate our analysis, we cite a special rank-one matrix decomposition lemma as follows.

\begin{lemma}[Theorem 2.1 in \cite{h-z05}]\label{dcmp2} Suppose that $\bX$ is an $N \times N$ complex Hermitian PSD matrix of rank $R$, and
$\bA$ and $\bB$ are given $N \times N$ Hermitian matrices. Then,
there is a rank-one decomposition
$\bX=\sum_{r=1}^R{\bx_r}{\bx_r^H}$ such that
\[
{\bx_r^H}{\bA}{\bx_r}=\frac{\tr(\bA\bX)}{R} \,\,\, \mbox{ and
} \,\,\, {\bx_r^H}{\bB}{\bx_r}=\frac{\tr(\bB\bX)}{R}, \,
r=1,\ldots,R.
\]
\end{lemma}

The rank-one decomposition synthetically is denoted as $\bx=\mathcal{D}_1(\bX,\bA,\bB)$.

Suppose that $y_1+y_2+y_3<0$. We have the following theorem.

\begin{theorem}\label{suffi-conditions-2}
Suppose that $(\bW,y_1,y_2,y_3,y_4)$ and $(\bZ,\bz_1,z_0,x)$ are the solutions for  primal SDP \eqref{rob-beamf-QMI-1-LMI-restricted} and  dual SDP \eqref{rob-beamf-QMI-1-LMI-restricted-dual}, respectively. Suppose that $\rank(\bW)$ is higher than one and $y=y_1+y_2+y_3<0$.
Suppose also that
%\begin{equation}\label{suffi-conditions-1-assumption-1}
%\tr(\bW)-\|\bar\ba\|^2<0,
%\end{equation}
%and
\begin{equation}\label{suffi-conditions-1-assumption-2}
y^2+y(\|\bar\ba\|^2-\tr\bW)+\bar\ba^H\bW\bar\ba-\|\bar\ba\|^2\tr\bW\ge0.
\end{equation}
Then there is a rank-one optimal solution for  SDP problem \eqref{rob-beamf-QMI-1-LMI-restricted}.
\end{theorem}
\proof
See Appendix \ref{proof-suffi-conditions-2}.
\endproof

Suppose $y_1+y_2+y_3=0$. Equation \eqref{LMI-constraint-reformulate-Schur-complement} reduces to
\[
\bW\succeq\bar\ba\bar\ba^H.
\]
Therefore, we have $\tr(\hat\bR\bW)\ge\bar\ba^H\hat\bR\bar\ba$.
It is known that if $\bw$ satisfies
\[
\bw\bw^H\succeq\bar\ba\bar\ba^H,
\]
then there is  $\lambda\in\mathbb{C}$ with $|\lambda|\ge1$ such that $\bw=\lambda\bar\ba$, and vice versa. Hence, the optimal value is equal to $|\lambda|^2\bar\ba^H\hat\bR\bar\ba$, provided that $\bw$ is an optimal solution. Observe that our goal is to minimize $\bw^H\hat\bR\bw$, and therefore, the optimal value is $\bar\ba^H\hat\bR\bar\ba$ when $|\lambda|=1$.
Then we obtain the following proposition.

\begin{proposition}\label{rnk1-soln-property}
Suppose that $(\bW,y_1,y_2,y_3,y_4)$ and $(\bZ,\bz_1,z_0,x)$ are the solutions for primal SDP \eqref{rob-beamf-QMI-1-LMI-restricted} and dual SDP \eqref{rob-beamf-QMI-1-LMI-restricted-dual}, respectively. Suppose that $\rank(\bW)$ is higher than one and $y_1+y_2+y_3=0$. %and $\bW=\bw\bw^H$ is of rank one.
Then $\bw=\bar\ba$ is optimal for \eqref{rob-beamf-QMI-restricted}, and thus, for \eqref{rob-beamf-QMI}.
\end{proposition}

%The following theorem states how to obtain an optimal solution from a high-rank solution $\bW$.
%\begin{theorem}\label{y1+y2+y3-equal-0}
%Suppose that $(\bW,y_1,y_2,y_3,y_4)$ and $(\bZ,\bz_1,z_0,x)$ are the solutions for the primal SDP \eqref{rob-beamf-QMI-1-LMI-restricted} and the dual SDP \eqref{rob-beamf-QMI-1-LMI-restricted-dual}, respectively. Suppose that $y_1+y_2+y_3=0$. Suppose that there is $\bW=\sum_{r=1}^R\bw_r\bw_r^H$ such that $\tr(\hat\bR\bW)=\tr(\hat\bR(\sqrt{R}\bw_r)(\sqrt{R}\bw_r)^H)$ for $r=1,\ldots,R$ (via Lemma \ref{dcmp2}); %suppose that there is another $\bW=\sum_{r=1}^R\bu_r\bu_r^H$ such that $\tr(\hat\bR\bW)=\tr(\hat\bR(\sqrt{R}\bu_r)(\sqrt{R}\bu_r)^H)$ and $\tr(\bW)=\tr((\sqrt{R}\bu_r)(\sqrt{R}\bu_r)^H)$, for $r=1,\ldots, R$.
% If a vector $\bw\in\{\sqrt{R}\bw_1,\ldots,\sqrt{R}\bw_R\}$ and a complex number $\lambda\in\mathbb{C}$ with $|\lambda|=1$ are such that $\bw=\lambda\bar\ba$, then $\bw\bw^H$ is a rank-one solution for SDP \eqref{rob-beamf-QMI-1-LMI-restricted}.
%\end{theorem}

\subsection{BLMI Approximation Approach}\label{BLMI-1}

Note that if the optimal solution $\bW^\star$ of \eqref{rob-beamf-QMI-1-LMI-restricted} is of rank one, then we claim that the solution $\bw^\star$ with $\bW^\star=\bw^\star\bw^{\star H}$ is optimal for \eqref{rob-beamf-QMI-restricted}, and thus, for \eqref{rob-beamf-QMI}. If the solution $\bW^\star$ is of higher rank, we need to build another method to solve \eqref{rob-beamf-QMI}.

%Suppose that a high-rank optimal solution $\bW_0:=\bW^\star$ for \eqref{rob-beamf-QMI-1-LMI} is obtained. In this regard, we prefer to find a rank-one optimal/suboptimal solution.
In this subsection, we present a BLMI reformulation for QMI problem \eqref{rob-beamf-QMI} and propose an approximation algorithm for such BLMI problem.% in order to solve \eqref{rob-beamf-QMI}.

Note that for a nonzero $\bW\succeq\bzero$, the statement that $\bW$ is of rank one is equivalent to that $\lambda_2(\bW)\le0$ (recall that throughout the paper, we place the eigenvalues of a matrix in a descending order, e.g., $\lambda_1(\bW)\ge\lambda_2(\bW)\ge\cdots\ge\lambda_N(\bW)$). Note that an optimal solution for the relaxation problem \eqref{rob-beamf-QMI-1-LMI-restricted} cannot be zero. In fact, if an optimal solution for \eqref{rob-beamf-QMI-1-LMI-restricted} is the zero matrix, then $\bw=\bzero$ is optimal for \eqref{rob-beamf-QMI-restricted}, and thus, for \eqref{rob-beamf-QMI}, which is a contradiction since we maximize the worst-case SINR  (it is always assumed to be positive and finite).

Clearly, we have that for the nonzero PSD matrix $\bW$, the following statement is true.
\[
\rank(\bW)=1\Leftrightarrow\lambda_2(\bW)\le0\Leftrightarrow \lambda_1(\bW)+\lambda_2(\bW)\le\lambda_1(\bW).
\]
Therefore, we can put the new rank-one constraint $\lambda_1(\bW)+\lambda_2(\bW)\le\lambda_1(\bW)$ into \eqref{rob-beamf-QMI-1-LMI}, and obtain a reformulation for \eqref{rob-beamf-QMI}, that is,
\begin{equation}\label{rob-beamf-QMI-1-LMI-1}%\mbox{(problem name)}%\left\{
\begin{array}[c]{ll}
\underset{\bW,\,\{y_i\}_{i=1}^4}{\sf{minimize}}  & \begin{array}[c]{c} \tr(\hat\bR\bW) \end{array}\\
\sf{subject\;to}               &
\begin{array}[t]{l}
\eqref{rob-beamf-QMI-1-LMI.b},\eqref{rob-beamf-QMI-1-LMI.c},\eqref{rob-beamf-QMI-1-LMI.d}\mbox{ satisfied},\\
%y_1+y_2+y_3\le0,\\
\lambda_1(\bW)+\lambda_2(\bW)\le\lambda_1(\bW).
\end{array}
\end{array}%\right.
\end{equation}

Note that in \eqref{rob-beamf-QMI-1-LMI-1}, the additional linear constraint \eqref{additional-linear-constr} is not included (unlike problem \eqref{rob-beamf-QMI-restricted} or \eqref{rob-beamf-QMI-1-LMI-restricted}), since the additional constraint is not needed any longer when we have the rank-one constraint in \eqref{rob-beamf-QMI-1-LMI-1}. It is worth highlighting  that this method is universal, namely, the constraint $\lambda_1(\bW)+\lambda_2(\bW)\le\lambda_1(\bW)$ can be added to an LMI relaxation problem for which one seeks a nonzero rank-one solution.

In order to represent the constraint $\lambda_1(\bW)+\lambda_2(\bW)\le\lambda_1(\bW)$, we invoke a result in \cite[Sec. 4.2, Example 18c]{Nemi-book2001}, which states the equivalence between the sum of $K$ largest eigenvalues of a real symmetric matrix and a finite LMI representation. For the completeness, we cite the result %(an extended version to the Hermitian matrix case)
as follows.

\begin{lemma}\label{nemirovski-result}
Suppose that $S_K(\bW)$ is the sum of $K$ largest eigenvalues of the Hermitian matrix $\bW$ ($K\le N$). Then the epigraph $\{(\bW,t)~|~S_K(\bW)\le t\}$ of the function admits the LMI representation
\begin{equation}\label{LMI-representation-1}
t-Ks-\tr(\bZ)\ge0,\,\bZ-\bW+s\bI\succeq\bzero,\,\bZ\succeq\bzero,
\end{equation}
where $\bZ$ is an $N\times N$ Hermitian matrix and $s$ is a real number.
\end{lemma}

We note that the original result \cite{Nemi-book2001} holds for a symmetric matrix $\bW$, but it can be extended to a Hermitian matrix straightforwardly, as already stated in the previous lemma.

Therefore, it follows from Lemma~\ref{nemirovski-result} that the constraint $\lambda_1(\bW)+\lambda_2(\bW)\le \lambda_1(\bW)$ has the following LMI representation.
\begin{equation}\label{sum-eigenv-sdr}
\lambda_1(\bW)-2s-\tr(\bZ)\ge0,\, \bZ-\bW+s\bI\succeq\bzero,\, \bZ\succeq\bzero,
\end{equation}
where $\bZ$ and $s$ are additional variables. Considering that
\begin{equation}\nonumber
\lambda_1(\bW)={\sf{maximize}}~\tr(\bW\bX)~~ {\sf{subject\;to}}~ \tr\bX=1,\,\bX\succeq\bzero,
\end{equation}
we can reexpress QMI problem \eqref{rob-beamf-QMI-1-LMI-1} into the following problem
\begin{equation}\label{rob-beamf-QMI-1-LMI-more}%\mbox{(problem name)}%\left\{
\begin{array}[c]{cl}
\underset{\bW,\bX,\bZ,\{y_i\}_{i=1}^4,s}{\sf{minimize}}  & \begin{array}[c]{c} \tr(\hat\bR\bW) \end{array}\\
\sf{subject\;to}               &
\begin{array}[t]{l}
\eqref{rob-beamf-QMI-1-LMI.b},\eqref{rob-beamf-QMI-1-LMI.c},\eqref{rob-beamf-QMI-1-LMI.d}\mbox{ satisfied},\\
% y_1+y_2+y_3\le0,\\
 \tr(\bW\bX)-2s-\tr(\bZ)\ge0,\\
 \bZ-\bW+s\bI\succeq\bzero,\\
 \tr\bX=1,\\
 \bZ\succeq\bzero,\,\bX\succeq\bzero,\,s\in\mathbb{R}.
\end{array}
\end{array}%\right.
\end{equation}

Suppose that $\bW^\star$ is the optimal solution of \eqref{rob-beamf-QMI-1-LMI-more}. %Clearly, if the optimal value $\tr(\hat\bR\bW^\star)$ of \eqref{rob-beamf-QMI-1-LMI-more} is equal to that of \eqref{rob-beamf-QMI-1-LMI}, then $\bW^\star$ must be an optimal solution for \eqref{rob-beamf-QMI-1-LMI}.
It is noticed that $\bW^\star$ must be a rank-one since $\lambda_2(\bW^\star)\le0$ (or $\lambda_1(\bW^\star)+\lambda_2(\bW^\star)\le\lambda_1(\bW^\star)$) has been fulfilled automatically in \eqref{rob-beamf-QMI-1-LMI-more}. Therefore, the vector $\bw^\star$ with $\bW^\star=\bw^\star\bw^{\star H}$ is optimal for \eqref{rob-beamf-QMI}, and thus, for original RAB problem \eqref{max-min-2003-2} with ${\cal A}={\cal A}_1$. %Clearly, if the optimal value $\tr(\hat\bR\bW^\star)$ of \eqref{rob-beamf-QMI-1-LMI-more} is equal to that of \eqref{rob-beamf-QMI-1-LMI}, then $\bW^\star$ must be an optimal solution for \eqref{rob-beamf-QMI-1-LMI-more}.

We observe that the only difficulty in \eqref{rob-beamf-QMI-1-LMI-more} is the bilinear term $\tr(\bW\bX)$ in the fifth constraint (for this reason, the problem is called a BLMI problem). To overcome the aforementioned difficulty with \eqref{rob-beamf-QMI-1-LMI-more}, we take the high-rank solution $\bW^\star$ for \eqref{rob-beamf-QMI-1-LMI-restricted} as an initial point $\bW_0$, and solve LMI problem \eqref{rob-beamf-QMI-1-LMI-more} with the fifth  constraint changed to $\tr(\bW_0\bX)-2s-\tr(\bZ)\ge0$, obtaining $\bW_1$. Set $\bW_0=\bW_1$ and repeat the step of solving \eqref{rob-beamf-QMI-1-LMI-more} with the fifth constraint changed, obtaining $\bW_2$. Repeat the previous two steps, until a stopping criteria (for example, $\|\bW_k-\bW_{k-1}\|_2\le \xi$, or $|\lambda_1(\bW_{k})-\lambda_1(\bW_{k-1})|\le\xi$) is satisfied. We summarize the procedure for solving \eqref{rob-beamf-QMI} in Algorithm~\ref{alg-1}.

\begin{algorithm}\caption{Procedure for Solving Problem \eqref{rob-beamf-QMI}}\label{alg-1}
\begin{algorithmic}[1]
\REQUIRE  $\hat\bR$, $\hat\ba$, $\eta_1$, $\eta_2$, $\epsilon$, $\xi$; %$\unrhd_m$,

\ENSURE A solution $\bw^\star$ for problem \eqref{rob-beamf-QMI}; %%Aqui van els outputs

\STATE solve \eqref{rob-beamf-QMI-1-LMI-restricted}, returning $\bW^\star$;

\IF {$\bW^\star$ is of rank one}

\STATE output  $\bw^\star$ with $\bW^\star=\bw^\star\bw^{\star H}$, and terminate;

\ENDIF

\STATE let $k=0$; let $\bW_0$ be the optimal solution $\bW^\star$ for  \eqref{rob-beamf-QMI-1-LMI-restricted};%, and set $\bW_k=\bW_0$;

\REPEAT

\STATE solve \eqref{rob-beamf-QMI-1-LMI-more} with the fifth constraint changed to $\tr(\bW_k\bX)-2s-\tr(\bZ)\ge0$, obtaining the solution $\bW_{k+1}$;

\STATE $k:=k+1$;

\UNTIL{$\|\bW_k-\bW_{k-1}\|_2\le \xi$}

%\STATE $\bW_k=\lambda_1\bw_1\bw_1^H+\cdots+\lambda_R\bw_R\bw_R^H$ (where $\lambda_1\ge\cdots\ge\lambda_R$);

\STATE output $\bw^\star=\sqrt{\lambda_1}\bw_1$, where $\lambda_1$ is the largest eigenvalue of $\bW_k$ and $\bw_1$ is a corresponding eigenvector.

\end{algorithmic}
\end{algorithm}

If the rank of $\bW_k$ in Step 10 of Algorithm~\ref{alg-1} is one, then $\bw^\star$ with $\bW^\star=\bw^\star\bw^{\star H}$ is the solution for  \eqref{rob-beamf-QMI}. Otherwise, the solution is $\bw^\star=\sqrt{\lambda_1}\bw_1$ as stated in Step 10. The computational cost of Algorithm \ref{alg-1} is dominated by solving the SDP problem in Step 7 of each iteration.

Note that it is possible that LMI problem \eqref{rob-beamf-QMI-1-LMI-more} with the fifth constraint changed to $\tr(\bW_k\bX)-2s-\tr(\bZ)\ge0$ leads to infeasibility due to several constraints  added. In this case, we set $\bW_k:=2\bW_k$ and repeat Step 7. By doing so, we wish to provide a way to enlarge the feasible set of \eqref{rob-beamf-QMI-1-LMI-more}. Our simulations show that a rank-one solution $\bW^\star:=\bW_k$ in Step  10  can always be obtained.%, which means that the procedure is convergent indeed.%, and it is workable in our simulations.

\section{Solving Robust Optimization Problem \eqref{rob-beamf} with Uncertainty Set ${\cal A}_2$ in \eqref{uncertainty-set-KVH2012}}

In this section, we study robust adaptive beamforming problem via worst-case SINR maximization with nonconvex uncertainty set ${\cal A}_2$ and its extensions.

\subsection{Robust Optimization Problem \eqref{rob-beamf} with Uncertainty Set ${\cal A}_2$} \label{sec-uncertainty-set-A2}
Let us first look into the minimization problem in the constraint of robust optimization problem \eqref{rob-beamf}, when ${\cal A} = {\cal A}_2$. The problem then can be written as
\begin{equation}\label{min-constraint-A2}%\mbox{(problem name)}%\left\{
\begin{array}[c]{ll}
\underset{\ba\in\mathbb{C}^N}{\sf{minimize}}  & \begin{array}[c]{c}\ba^H\bw\bw^H\ba\end{array}\\
\sf{subject\;to}               &
\begin{array}[t]{l}
\ba^H\bar\bC\ba\le \Delta_0,\\
N-\eta_1\le\|\ba\|^2\le N+\eta_2.
\end{array}
\end{array}%\right.
\end{equation}

%Suppose that the eigenvalues $\{\lambda_n(\bar\bC)\}$ of $\bar\bC$ are placed in a decreasing order, namely, $\lambda_1(\bar\bC)\ge\ldots\ge\lambda_N(\bar\bC)$.
From the feasibility of \eqref{min-constraint-A2}, we have
\[
\lambda_N(\bar\bC)\le\frac{\ba^H\bar\bC\ba}{\|\ba\|^2}\le\frac{\Delta_0}{N-\eta_1};
\]
in other words,
\begin{equation}\label{delta0-lowerbound}
\lambda_N(\bar\bC)(N-\eta_1)\le\Delta_0.
\end{equation}
On the other hand, if $\Delta_0$ is sufficiently large, then the constraint $\ba^H\bar\bC\ba\le \Delta_0$ vanishes in the feasible set since it is fulfilled always. We claim that if $\Delta_0\ge\lambda_1(\bar\bC)(N+\eta_2)$, then it follows that $\ba^H\bar\bC\ba\le \Delta_0$ always holds. In fact, $\Delta_0\ge\lambda_1(\bar\bC)(N+\eta_2)\ge\lambda_1(\bar\bC)\|\ba\|^2$, which means that $\Delta_0/\|\ba\|^2\ge\lambda_1(\bar\bC)\ge\ba^H\bar\bC \ba/ \| \ba \|^2$. In other words, $\Delta_0\ge\ba^H\bar\bC\ba$. Therefore, in order to make the first inequality constraint in \eqref{min-constraint-A2} hold, it should be satisfied that
\begin{equation}\label{delta0-lowerbound-upperbound}
\lambda_N(\bar\bC)(N-\eta_1)\le\Delta_0\le\lambda_1(\bar\bC)(N+\eta_2),
\end{equation}
considering \eqref{delta0-lowerbound}.

Suppose that there is a vector $\ba_0$ such that $\ba_0^H\bar\bC\ba_0<\Delta_0$ and $\|\ba_0\|^2=N$. In other words, problem \eqref{min-constraint-A2} is strictly feasible.

The SDP relaxation for \eqref{min-constraint-A2} is
\begin{equation}\label{min-constraint-A2-SDR}%\mbox{(problem name)}%\left\{
\begin{array}[c]{ll}
\underset{\bX\in{\cal{H}}^N}{\sf{minimize}}  & \begin{array}[c]{c}\tr(\bw\bw^H\bX)\end{array}\\
\sf{subject\;to}               &
\begin{array}[t]{l}
\tr(\bar\bC\bX)\le \Delta_0,\\
N-\eta_1\le \tr\bX\le N+\eta_2,\\
\bX\succeq\bzero
\end{array}
\end{array}%\right.
\end{equation}
and its dual problem can be written as
\begin{equation}\label{min-constraint-A2-SDR-dual}%\mbox{(problem name)}%\left\{
\begin{array}[c]{ll}
\underset{x,y_1,y_2}{\sf{maximize}}  & \begin{array}[c]{c}\Delta_0x+(N-\eta_1)y_1+(N+\eta_2)y_2\end{array}\\
\sf{subject\;to}               &
\begin{array}[t]{l}
\bw\bw^H-x\bar\bC-(y_1+y_2)\bI\succeq\bzero,\\
x\le0,\,y_1\ge0,\,y_2\le0.
\end{array}
\end{array}%\right.
\end{equation}
It can be seen easily that the dual SDP is strictly feasible. The primal SDP is also strictly feasible since the point $\epsilon\bI+(1-\epsilon)\ba_0\ba_0^H$ is a strictly feasible point for a sufficiently small $\epsilon>0$. Therefore, it follows again from the strong duality theorem %(see e.g. \cite[Theorem 1.4.2]{Nemirovski})
that problems \eqref{min-constraint-A2-SDR} and \eqref{min-constraint-A2-SDR-dual} are solvable and there is zero gap between them. In addition, there are only two constraints in the primal SDP, and then a rank-one solution for it can be always constructed efficiently (see, e.g., \cite{h-z05}). Thereby, SDP relaxation \eqref{min-constraint-A2-SDR} is tight, and we have $v^\star(\eqref{min-constraint-A2})=v^\star(\eqref{min-constraint-A2-SDR})=v^\star(\eqref{min-constraint-A2-SDR-dual})$.

Therefore, robust beamforming problem \eqref{rob-beamf} with  ${\cal A}_2$ can be recast into the following QMI problem:
\begin{subequations}\label{rob-beamf-QMI-A2}%\mbox{(problem name)}%\left\{
\begin{align}\label{rob-beamf-QMI-A2.a}
\underset{\bw,x,y_1,y_2}{\sf{minimize}}  & \quad\bw^H\hat\bR\bw \\ \label{rob-beamf-QMI-A2.b}
\sf{subject\;to}               & \quad\Delta_0x+(N-\eta_1)y_1+(N+\eta_2)y_2=1,\\ \label{rob-beamf-QMI-A2.c}
%\bA_0-y_1\bA_1-(y_2+y_3)\bA_2-y_4\bA_3\succeq\bzero,\\
                               & \quad\bw\bw^H-x\bar\bC-(y_1+y_2)\bI\succeq\bzero,\\ \label{rob-beamf-QMI-A2.d}
                               & \quad x\le0,\,y_1\ge0,\,y_2\le0.
\end{align}%\right.
\end{subequations}
Its LMI relaxation can be written as
\begin{subequations}\label{rob-beamf-QMI-A2-SDR}%\mbox{(problem name)}%\left\{
\begin{align}\label{rob-beamf-QMI-A2-SDR.a}
\underset{\bW,x,y_1,y_2}{\sf{minimize}}  & \quad \begin{array}[c]{c}\tr(\hat\bR\bW) \end{array}\\ \label{rob-beamf-QMI-A2-SDR.b}
\sf{subject\;to}               & \quad \Delta_0x+(N-\eta_1)y_1+(N+\eta_2)y_2=1,\\  \label{rob-beamf-QMI-A2-SDR.c}
%\bA_0-y_1\bA_1-(y_2+y_3)\bA_2-y_4\bA_3\succeq\bzero,\\
 & \quad \bW-x\bar\bC-(y_1+y_2)\bI\succeq\bzero,\\ \label{rob-beamf-QMI-A2-SDR.d}
 & \quad \bW\succeq\bzero, x\le0,\,y_1\ge0,\,y_2\le0.
\end{align}%\right.
\end{subequations}
%By the way, it is easily verified that the dual problem is
%\begin{equation}\label{rob-beamf-QMI-A2-dual}%\mbox{(problem name)}%\left\{
%\begin{array}[c]{ll}
%\underset{\bZ,\,z}{\sf{maximize}}  & \begin{array}[c]{c} z \end{array}\\
%\sf{subject\;to}               &
%\begin{array}[t]{l}
%\hat\bR-\bZ\succeq\bzero,\\
%\tr(\bar\bC\bZ)-uz\le0,\\
%\tr\bZ-(N-\eta_1)z\ge0,\\
% \tr\bZ-(N+\eta_2)z\le0,\\
%\bZ\succeq\bzero,\,z\in\mathbb{R}.
%\end{array}
%\end{array}%\right.
%\end{equation}
If the solution of LMI problem \eqref{rob-beamf-QMI-A2-SDR} is of rank one, then problem \eqref{rob-beamf-QMI-A2} is solved. Otherwise, we consider a tightened LMI relaxation by adding one more valid linear constraint, applying the similar idea demonstrated in the previous section.

\subsection{Tightened LMI Relaxation}

The second constraint of \eqref{rob-beamf-QMI-A2} implies
\begin{equation}\label{additional-linear-constraints-A2}
x\lambda_{N-1}(\bar\bC)+y_1+y_2\le0,
\end{equation} due to $x\le0$.
Therefore, putting the additional linear constraint \eqref{additional-linear-constraints-A2} into \eqref{rob-beamf-QMI-A2} does not change the set of optimal solutions. Namely, the following problem
\begin{equation}\label{rob-beamf-QMI-A2-additonal-linear-constraint}%\mbox{(problem name)}%\left\{
\begin{array}[c]{ll}
\underset{\bw,x,y_1,y_2}{\sf{minimize}}  & \begin{array}[c]{c}\bw^H\hat\bR\bw \end{array}\\
\sf{subject\;to}               &
\begin{array}[t]{l}\eqref{rob-beamf-QMI-A2.b},\eqref{rob-beamf-QMI-A2.c},\eqref{rob-beamf-QMI-A2.d},\eqref{additional-linear-constraints-A2}\mbox{ satisfied},%\\
%x\lambda_{N-1}(\bar\bC)+y_1+y_2\le0,
\end{array}
\end{array}%\right.
\end{equation}
shares the same optimal solution set and the same optimal value with problem \eqref{rob-beamf-QMI-A2}.
Thus, we obtain the following restricted LMI relaxation problem
\begin{equation}\label{rob-beamf-QMI-A2-LMI-res}%\mbox{(problem name)}%\left\{
\begin{array}[c]{ll}
\underset{\bW,x,y_1,y_2}{\sf{minimize}}  & \begin{array}[c]{c}\tr(\hat\bR\bW) \end{array}\\
\sf{subject\;to}               &
\begin{array}[t]{l}\eqref{rob-beamf-QMI-A2-SDR.b},\eqref{rob-beamf-QMI-A2-SDR.c},\eqref{rob-beamf-QMI-A2-SDR.d},\eqref{additional-linear-constraints-A2}\mbox{ satisfied}.%\\
%x\lambda_{N-1}(\bar\bC)+y_1+y_2\le0.
\end{array}
\end{array}%\right.
\end{equation}
Hence, if rank-one solution $\bw^\star\bw^{\star H}$ is optimal for \eqref{rob-beamf-QMI-A2-LMI-res}, then the vector $\bw^\star$ is optimal for \eqref{rob-beamf-QMI-A2-additonal-linear-constraint}, and thus, for \eqref{rob-beamf-QMI-A2}.
%Here we note that only two additional constraints in \eqref{additional-linear-constraints-A2} are added due to the monotonicity the eigenvalues.

Clearly, the dual of LMI problem \eqref{rob-beamf-QMI-A2-LMI-res} is
\begin{equation}\label{rob-beamf-QMI-A2-LMI-res-dual}%\mbox{(problem name)}%\left\{
\begin{array}[c]{ll}
\underset{\bZ,\,z_1,\,z_2}{\sf{maximize}}  & \begin{array}[c]{c}z_1\end{array}\\
\sf{subject\;to}
& \begin{array}[t]{l}
\hat\bR-\bZ\succeq\bzero,\\
\tr(\bar\bC\bZ)-\Delta_0 z_1+\lambda_{N-1}(\bar\bC)z_2\le0,\\
\tr\bZ+z_2\le z_1(N+\eta_2),\\
\tr\bZ+z_2\ge z_1(N-\eta_1),\\
\bZ\succeq\bzero,\,z_1\in\mathbb{R},\,z_2\ge0.
\end{array}
\end{array}%\right.
\end{equation}
Suppose that the primal and dual SDPs are solvable, and possess the same optimal value (assumed to be positive and finite). Thus, the complementary conditions include
\begin{subequations}
\begin{align}\label{QMI-2-complementary-conditions-1}
\tr((\hat\bR-\bZ)\bW)&=0,\\ \label{QMI-2-complementary-conditions-2}
x(\tr(\bar\bC\bZ)-\Delta_0 z_1+\lambda_{N-1}(\bar\bC)z_2)&=0,\\ \label{QMI-2-complementary-conditions-3}
y_1(\tr\bZ+z_2-(N-\eta_1)z_1)&=0, \\ \label{QMI-2-complementary-conditions-4}
y_2(\tr\bZ+z_2-(N+\eta_2)z_1)&=0, \\ \label{QMI-2-complementary-conditions-5}
z_2(x\lambda_{N-1}(\bar\bC)+y_1+y_2) &=0,\\ \label{QMI-2-complementary-conditions-6}
 \tr(\bZ(\bW-x\bar\bC-(y_1+y_2)\bI))&=0.
\end{align}
\end{subequations}
It follows that $\tr(\hat\bR\bW)=z_1$ (i.e., the duality gap is zero). Based on the complementary conditions, we claim some necessary optimality conditions for \eqref{rob-beamf-QMI-A2-LMI-res} and \eqref{rob-beamf-QMI-A2-LMI-res-dual}.

\begin{proposition}\label{2-solution-property-1}
Suppose that $(\bW,x,y_1,y_2)$ and $(\bZ,z_1,z_2)$ are the solutions for primal SDP \eqref{rob-beamf-QMI-A2-LMI-res} and dual SDP \eqref{rob-beamf-QMI-A2-LMI-res-dual}, respectively. If $x\lambda_{N-1}(\bar\bC)+y_1+y_2<0$, then it holds that
\begin{enumerate}
\item $x<0$;
\item $z_1=\frac{\tr(\bar\bC\bZ)}{\Delta_0}$ and $z_2=0$;
\item $y_1+y_2>0$.
\end{enumerate}
\end{proposition}
\proof
See Appendix \ref{proof-2-solution-property-1}.
\endproof

\begin{proposition} \label{2-solution-property-2}
Suppose that $(\bW,x,y_1,y_2)$ and $(\bZ,z_1,z_2)$ are the solutions for primal SDP \eqref{rob-beamf-QMI-A2-LMI-res} and dual SDP \eqref{rob-beamf-QMI-A2-LMI-res-dual}, respectively. If $x\lambda_{N-1}(\bar\bC)+y_1+y_2=0$, then the following statements are true:
\begin{enumerate}
\item $x<0$;
\item $z_1=\frac{\tr(\bar\bC\bZ)+z_2\lambda_{N-1}(\bar\bC)}{\Delta_0}$;
\item $y_1+y_2>0$;
\item $\lambda_1(\bW)\ge-x(\lambda_{N-1}(\bar\bC)-\lambda_N(\bar\bC))$;
\item if $\bW=\bw\bw^H$ and $\lambda_{N-1}(\bar\bC)>\lambda_N(\bar\bC)$, then the optimal value $\bw^H\hat\bR\bw\ge-x(\lambda_{N-1}(\bar\bC)-\lambda_N(\bar\bC))\lambda_N(\hat\bR)$. %$\bw^H\hat\bR\bw\ge-x(\lambda_{N-1}(\bar\bC)\lambda_N(\hat\bR)-\lambda_N(\hat\bR\bar\bC))$.
\end{enumerate}
\end{proposition}
\proof
See Appendix \ref{proof-2-solution-property-2}.
\endproof

%In particular, when the solution $\bW=\bw\bw^H$ is of rank-one, the fourth statement of Proposition \ref{2-solution-property-2} implies that $\|\bw\|^2$ has the lower bound $-x(\lambda_{N-1}(\bar\bC)-\lambda_N(\bar\bC))$, and when $\hat\bR=\bI$, the fifth statement means the same lower bound for $\|\bw\|^2$.

To establish a sufficient condition for \eqref{rob-beamf-QMI-A2-LMI-res} to have a rank-one solution, we give the following theorem.
\begin{theorem} \label{sufficient-conditions-3}
Suppose that $(\bW,x,y_1,y_2)$ and $(\bZ,z_1,z_2)$ are the solutions for primal SDP \eqref{rob-beamf-QMI-A2-LMI-res} and dual SDP \eqref{rob-beamf-QMI-A2-LMI-res-dual}, respectively. If $\tr(\hat\bR^{-1}\bZ)\le1$, then $\bW$ must be of rank one.
\end{theorem}
\proof
See Appendix \ref{proof-sufficient-conditions-3}.
\endproof

\subsection{BLMI Approximation Approach}

Suppose that the solution $\bW$ for \eqref{rob-beamf-QMI-A2-LMI-res} has rank higher than one. Therefore, we can apply the BLMI approximation method to obtain a solution for \eqref{rob-beamf-QMI-A2}.
Evidently, QMI problem \eqref{rob-beamf-QMI-A2} can be rewritten as
\begin{equation}\label{rob-beamf-QMI-A2-SDR-BLMI}%\mbox{(problem name)}%\left\{
\begin{array}[c]{ll}
\underset{\bW,x,y_1,y_2}{\sf{minimize}}  & \begin{array}[c]{c}\tr(\hat\bR\bW) \end{array}\\
\sf{subject\;to}               &
\begin{array}[t]{l}\eqref{rob-beamf-QMI-A2-SDR.b},\eqref{rob-beamf-QMI-A2-SDR.c},\eqref{rob-beamf-QMI-A2-SDR.d} \mbox{ satisfied}, \\
%x\lambda_{N-1}(\bar\bC)+y_1+y_2\le0,\\
\lambda_1(\bW)+\lambda_2(\bW)\le\lambda_1(\bW).
\end{array}
\end{array}%\right.
\end{equation}
Moreover, problem \eqref{rob-beamf-QMI-A2-SDR-BLMI} can be further recast into
\begin{equation}\label{rob-beamf-QMI-A2-SDR-BLMI-1}%\mbox{(problem name)}%\left\{
\begin{array}[c]{cl}
\underset{\bW,\bX, \bZ, x,y_1,y_2,s}{\sf{minimize}}  & \begin{array}[c]{c}\tr(\hat\bR\bW) \end{array}\\
\sf{subject\;to}               &
\begin{array}[t]{l}\eqref{rob-beamf-QMI-A2-SDR.b},\eqref{rob-beamf-QMI-A2-SDR.c},\eqref{rob-beamf-QMI-A2-SDR.d} \mbox{ satisfied}, \\
%x\lambda_{N-1}(\bar\bC)+y_1+y_2\le0,\\
\tr(\bW\bX)-2s-\tr(\bZ)\ge0,\\
\bZ-\bW+s\bI\succeq\bzero,\\
\tr\bX=1,\\
\bZ\succeq\bzero,\,\bX\succeq\bzero,\,s\in\mathbb{R}.\\
\end{array}
\end{array}%\right.
\end{equation}
This problem can be solved in a similar way as problem \eqref{rob-beamf-QMI-1-LMI-more} is solved. %Precisely, take a high-rank solution $\bW^\star$ for \eqref{rob-beamf-QMI-A2-SDR} as an initial point $\bW_k$ with $k=0$, and then solve \eqref{rob-beamf-QMI-A2-SDR-BLMI-1} with the third constraint changed to $\tr(\bW_k\bX)-2s-\tr(\bZ)\ge0$, obtaining an solution $\bW_{k+1}$. Set $k:=k+1$. Repeat to solve \eqref{rob-beamf-QMI-A2-SDR-BLMI-1} with the third constraint changed. Repeat the previous two steps until a stopping criteria, say $\|\bW_{k}-\bW_{k-1}\|_2\le\xi$, is satisfied.
We summarize the procedure in Algorithm~\ref{alg-rob-beamf-QMI-A2-SDR}.

\begin{algorithm}\caption{Procedure for Solving Problem \eqref{rob-beamf-QMI-A2}}\label{alg-rob-beamf-QMI-A2-SDR}
\begin{algorithmic}[1]
\REQUIRE  $\hat\bR$, $\bar\bC$, $\Delta_0$, $\eta_1$, $\eta_2$, $\xi$; %$\unrhd_m$,

\ENSURE A solution $\bw^\star$ for problem \eqref{rob-beamf-QMI-A2}; %%Aqui van els outputs

\STATE solve \eqref{rob-beamf-QMI-A2-LMI-res}, returning $\bW^\star$;

\IF {$\bW^\star$ is of rank one}

\STATE output  $\bw^\star$ with $\bW^\star=\bw^\star\bw^{\star H}$, and terminate;

\ENDIF

\STATE let $k=0$; let $\bW_k$ be the optimal solution $\bW^\star$ for \eqref{rob-beamf-QMI-A2-LMI-res};

\REPEAT

\STATE solve \eqref{rob-beamf-QMI-A2-SDR-BLMI-1} with the fourth constraint changed to $\tr(\bW_k\bX)-2s-\tr(\bZ)\ge0$, obtaining the solution $\bW_{k+1}$;

\STATE $k:=k+1$;

\UNTIL{$\|\bW_k-\bW_{k-1}\|_2\le \xi$}

%\STATE $\bW_k=\lambda_1\bw_1\bw_1^H+\cdots+\lambda_R\bw_R\bw_R^H$ (where $\lambda_1\ge\cdots\ge\lambda_R$);

\STATE output $\bw^\star=\sqrt{\lambda_1}\bw_1$, where $\lambda_1$ is the largest eigenvalue of $\bW_k$ and $\bw_1$ is a corresponding eigenvector.

\end{algorithmic}
\end{algorithm}

If the solution in Step 10 $\bW_k=\bw^\star\bw^{\star H}$ is of rank one, then $\bw^\star$ is a solution for \eqref{rob-beamf-QMI-A2}. Otherwise, we regard $\sqrt{\lambda_1}\bw_1$ as a solution for \eqref{rob-beamf-QMI-A2}, where $\lambda_1$ is the largest eigenvalue and $\bw_1$ is a corresponding eigenvector. If the problem solved in Step 7 is infeasible, we adopt the same strategy as in Subsection~\ref{BLMI-1} to enlarge the feasible set and continue to solve it in an approximate way.

%{\color{cyan}In a similar way to Algorithm \ref{alg-1-1}, we can solve \eqref{rob-beamf-QMI-A2} by apply Algorithm \ref{alg-rob-beamf-QMI-A2-SDR} but with step 7 changed to solving the following problem:
%\begin{equation}\label{rob-beamf-QMI-A2-SDR-BLMI-1-lambda-1}%\mbox{(problem name)}%\left\{
%\begin{array}[c]{ll}
%\underset{}{\sf{minimize}}  & \begin{array}[c]{c}\tr(\hat\bR\bW) \end{array}\\
%\sf{subject\;to}               &
%\begin{array}[t]{l}\Delta_0x+(N-\eta_1)y_1+(N+\eta_2)y_2=1,\\
%\bW-x\bar\bC-(y_1+y_2)\bI\succeq\bzero,\\
%x\lambda_{N-1}(\bar\bC)+y_1+y_2\le0,\\
%\lambda_1(\bW_k)-2s-\tr(\bZ)\ge0,\\
%\bZ-\bW+s\bI\succeq\bzero,\\
%\bW\succeq\bzero,\,\bZ\succeq\bzero,\\
%\,x\le0,\,y_1\ge0,\,y_2\le0,\,s\in\mathbb{R},
%\end{array}
%\end{array}%\right.
%\end{equation}
%obtaining solution $\bW_{k+1}$.
%}

%\subsection{Robust optimization problem \eqref{max-min-2003-2} with the uncertainty set ${\cal A}_3$}%\label{sec.4.3}
In order to solve robust problem \eqref{max-min-2003-2} with the uncertainty set ${\cal A}_3$, we note that
\[
{\cal A}_3=\{\ba~|~\ba^H\bC\ba\ge\Delta_1,\,N-\eta_1\le\|\ba\|^2\le N+\eta_2\}
\]
is equivalent to
\begin{equation}\label{A3-equiv}
{\cal A}_3=\{\ba~|~\ba^H(-\bC)\ba\le-\Delta_1,\,N-\eta_1\le\|\ba\|^2\le N+\eta_2\}.
\end{equation}
In this case, the valid linear constraint is
\[
-x\lambda_2(\bC)+y_1+y_2\le0,
\]
which can be added to enhance the LMI relaxation problem.
Therefore, Algorithm~\ref{alg-rob-beamf-QMI-A2-SDR} can be applied to solve the problem as well.

By adding one more similarity constraint, the uncertainty set ${\cal A}_2$ can be generalized to the following set
\begin{equation}\label{uncertainty_set_A2prime_defined}
\begin{array}[t]{l}
{\cal A}_4=\{\ba~|~\ba^H\bar\bC\ba\le\Delta_0,\,N-\eta_1\le\|\ba\|^2\le N+\eta_2, \\
~~~~~~~~~~~~~\|\ba-\hat\ba\|^2\le\epsilon\},
\end{array}
\end{equation}
and the uncertainty set ${\cal A}_3$ can be extended as the following set
\begin{equation}\label{uncertainty_set_A3prime_defined}
\begin{array}[t]{l}
{\cal A}_5=\{\ba~|~\ba^H\bC\ba\ge\Delta_1,\,N-\eta_1\le\|\ba\|^2\le N+\eta_2, \\
~~~~~~~~~~~~~\|\ba-\hat\ba\|^2\le\epsilon\}.
\end{array}
\end{equation}
In \cite{HZV2019tsp}, MVDR RAB problems with ${\cal A}_4$ and ${\cal A}_5$ have been studied. However, robust beamforming problem \eqref{max-min-2003-2} with either ${\cal A}_4$ or ${\cal A}_5$ can be discussed as well, according to the same vein applied herein. Then, Algorithm \ref{alg-rob-beamf-QMI-A2-SDR} modified accordingly can also be employed to solve it.

\section{Numerical Examples}

Let us consider a uniform linear array with $N=12$ omni-directional sensors spaced half a wavelength appart from each other, and the array noise is a spatially and temporally white Gaussian vector with zero mean and covariance matrix $\bI$. Two interferers with the same interference-to-noise ratio (INR) of 30~dB are assumed to impinge upon the array from the angles $\theta_1=-15^\circ$ and $\theta_2=15^\circ$ with respect to the array broadside, and the desired  signal is always present in the training data cell. The training sample size $T$ is preset to $100$ snapshots. The signal of interest impinges upon the array from the direction $\theta=7^\circ$ while the presumed direction is $\theta_0=5^\circ$ (thus $\hat\ba = \bd(\theta_0)$). %{\color{red}, while the presumed direction is assumed to be $5^\circ$ (i.e. $\theta_0$).}
The norm perturbation parameters  $\eta_1$ and $\eta_2$ are both set to $0.2N$, and $\epsilon=0.3N$. All results are averaged over 200 simulation runs. Parameter $\gamma$ in \eqref{uncertainty-set-B1} satisfies $\sqrt{\gamma}=0.1\lambda_{N}(\hat\bR)$, where the data sample covariance matrix $\bar\bR$ is different in each run.

In addition to the signal look direction mismatch, we take into account mismatch caused also by wavefront distortion in an inhomogeneous
medium \cite{KVH2012tsp}. That is, we assume that the signal  steering vector is distorted by wave propagation effects in the way that independent-increment phase distortions are accumulated by the components of the steering vector. We assume that the phase increments are independent Gaussian variables each with zero mean and standard deviation 0.02, and they are randomly generated and remain unaltered in each simulation run.

\subsection{Example 1: Uncertainty Set ${\cal A}_1$ (i.e., \eqref{uncertainty-set})}

\begin{figure}[!h]
\centerline{\resizebox{.55\textwidth}{!}{\includegraphics{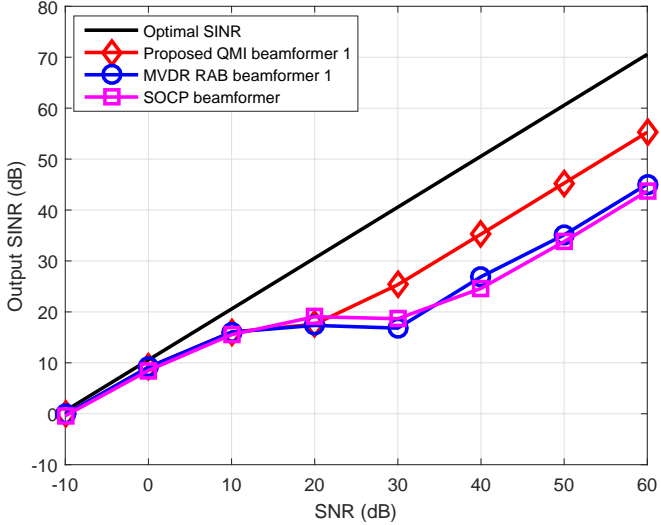}}
}
\vspace*{-.55\baselineskip}
\caption{Average beamformer output SINR versus SNR with $T=100$.}
\label{fig-sinr-snr}
\vspace*{0\baselineskip}
\end{figure}

\begin{figure}[!h]
\centerline{\resizebox{.55\textwidth}{!}{\includegraphics{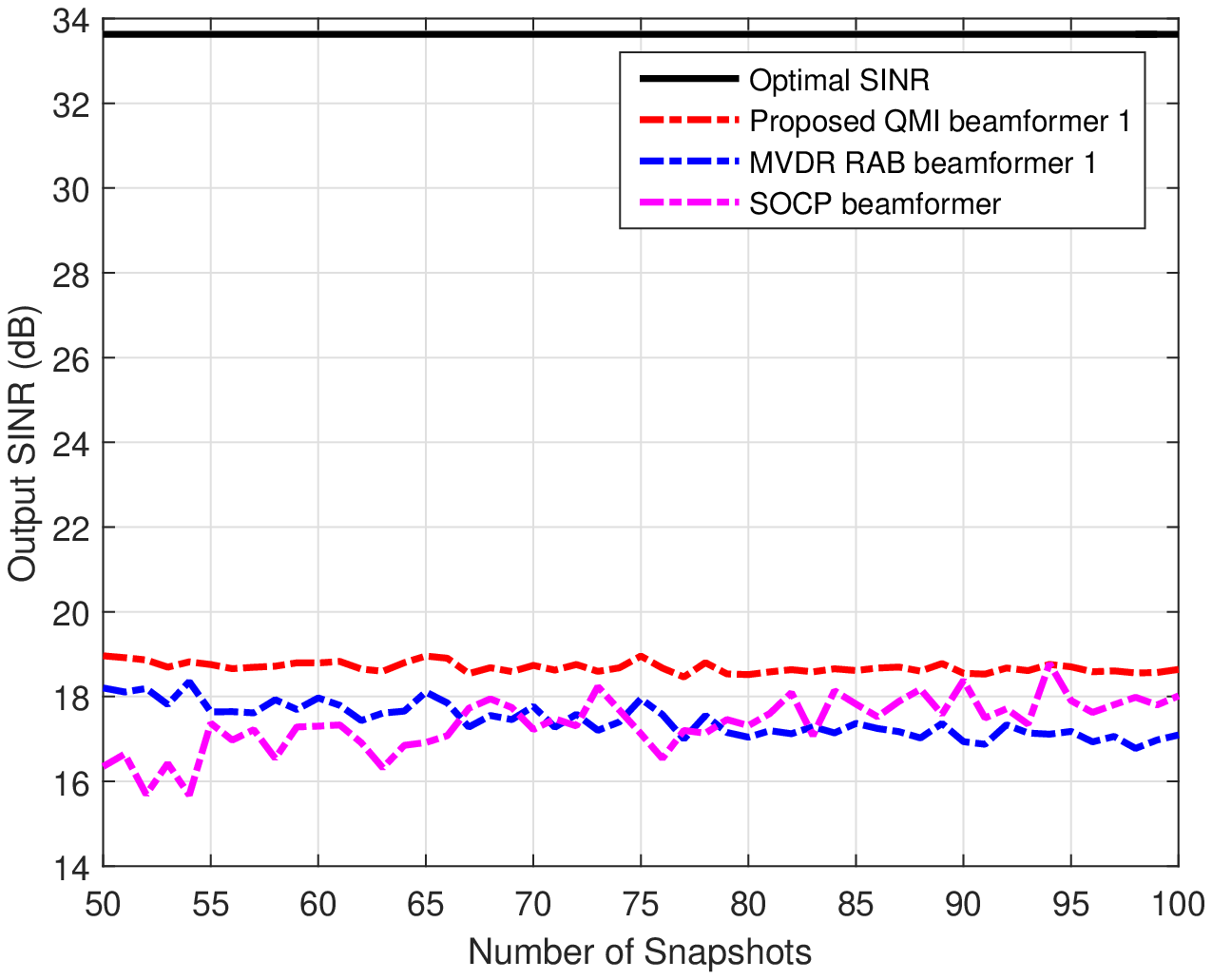}}
}
\vspace*{0.3\baselineskip}
\caption{Average array output power versus number of snapshots with SNR equal to 23~dB.}
\label{fig-sinr-snap}
\vspace*{0\baselineskip}
\end{figure}

In this example, the three problems, i.e., problem \eqref{rob-beamf-QMI} (via Algorithm \ref{alg-1}), problem \eqref{robust-opt-Vor03-socp-equiv} (also cf. \cite{VGL03tsp}) and MVDR RAB problem \eqref{opt-est-steering-vector-general} with uncertain set ${\cal A}_1$ (via, e.g., \cite[Algorithm 1]{HZV2019tsp}), are solved in every simulation run. The  beamformers obtained from the aforementioned three methods are termed respectively as ``Proposed QMI beamformer~1", ``SOCP beamformer", and ``MVDR RAB beamformer~1" in our figures.
Fig.~\ref{fig-sinr-snr} demonstrates the beamformer output SINR versus the  signal-to-noise ratio (SNR). % for the desired user.
As we can see, the output SINR obtained through \eqref{rob-beamf-QMI} is better than that obtained through the other two beamformers, especially at moderate and high SNRs. Fig.~\ref{fig-sinr-snap} depicts the output SINR versus the number of snapshots with SNR equal to 23~dB ($=10\log_{10}(200)$). From the figure, we observe that our proposed beamformer always provides higher SINR than the other two beamformers.

\subsection{Example 2: Uncertainty Sets ${\cal A}_2$ (i.e., \eqref{uncertainty-set-KVH2012}) and ${\cal A}_3$ (i.e., \eqref{uncertainty-set-HZV2019})}
\begin{figure}[!h]
\centerline{\resizebox{.55\textwidth}{!}{\includegraphics{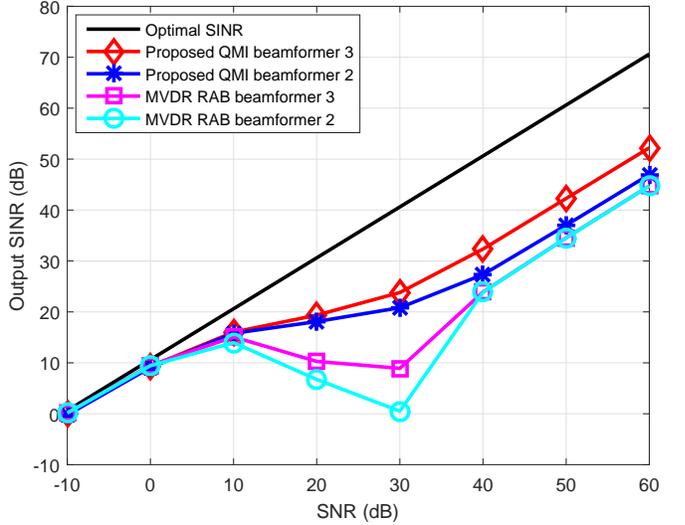}}
}
\vspace*{-.55\baselineskip}
\caption{Average beamformer output SINR versus SNR with $\Theta=[0^\circ,10^\circ]$ and $T=100$.}
\label{fig-sinr-snr-WC-KVH}
\vspace*{0\baselineskip}
\end{figure}

\begin{figure}[!h]
\centerline{\resizebox{.55\textwidth}{!}{\includegraphics{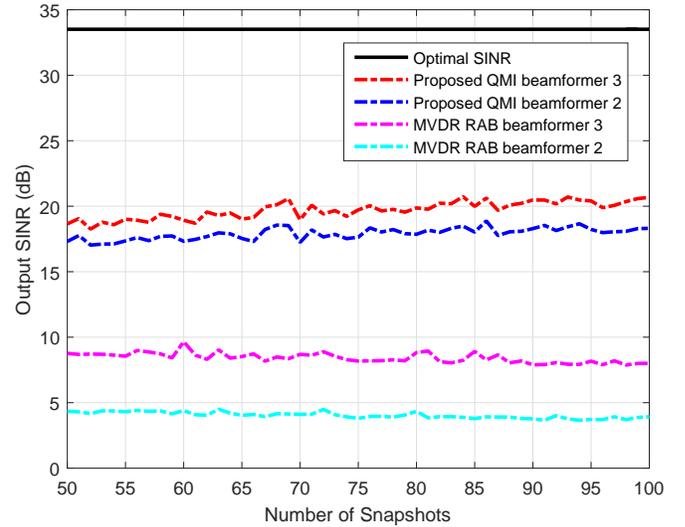}}
}
\vspace*{0.3\baselineskip}
\caption{Average array  output power versus number of snapshots with $\Theta=[0^\circ, 10^\circ]$ and SNR equal to 23~dB.}
\label{fig-sinr-snap-WC-KVH}
\vspace*{0\baselineskip}
\end{figure}

Suppose that the angular sector $\Theta$ of interest is $[0^\circ,10^\circ]$. Problem \eqref{max-min-2003-2} with ${\cal A}={\cal A}_2$ (i.e., QMI problem \eqref{rob-beamf-QMI-A2}, via Algorithm~\ref{alg-rob-beamf-QMI-A2-SDR}), and problem \eqref{max-min-2003-2} with ${\cal A}={\cal A}_3$ (i.e., QMI problem \eqref{rob-beamf-QMI-A2} with \eqref{A3-equiv}, via Algorithm~\ref{alg-rob-beamf-QMI-A2-SDR}) are solved, as well as MVDR RAB problems \eqref{opt-est-steering-vector-general} with ${\cal A}={\cal A}_2$ and ${\cal A}={\cal A}_3$ are solved via \cite[Algorithm 1]{HZV2019tsp} in a similar way. The corresponding beamformers from the previous four problems are called respectively, ``Proposed QMI beamformer~2", ``Proposed QMI beamformer~3", ``MVDR RAB beamformer~2", and ``MVDR RAB beamformer~3" in our figures. Fig.~\ref{fig-sinr-snr-WC-KVH} displays the array output SINR versus the SNR. We can observe from the figure that the output SINR through the worst-case SINR maximization based beamformer (the proposed QMI beamformer~2 or~3) is higher than that obtained through the MVDR RAB beamformer~2 or~3. Also, it can be seen that the beamformer (either the QMI or MVDR RAB beamformer) obtained by corresponding robust problem with ${\cal A}_3$ has better performance than the beamformers obtained by solving the robust problem with ${\cal A}_2$. Fig.~\ref{fig-sinr-snap-WC-KVH} shows the output SINR versus the number of snapshots when SNR$=23$~dB. From the figure, we can clearly confirm similar behaviors to those in Fig.~\ref{fig-sinr-snr-WC-KVH} for the four different beamformers.

\section{Conclusion}
We have considered the RAB problem by maximizing worst-case SINR with two types of nonconvex uncertainty sets for the steering vectors. We have reformulated the SINR maximization problem as QMI problems, and proposed tightened LMI relaxations for them. Some necessary and sufficient conditions for the LMI relaxation problems to admit a rank-one solution have been established. Then the LMI problems with the rank-one constraint have been further recast into corresponding BLMI problems, and an algorithm to solve such BLMI problems has been proposed, returning an optimal/suboptimal solution for the original RAB problem. The improved performance of the proposed robust beamformers has been demonstrated by simulations in terms of the array output SINR.

\appendix

\subsection{Proof of Lemma \ref{strict-feasibility-SDPs-primal-dual}}\label{strict-feasibility-SDPs-primal-dual-proof-appendix}
\proof
It can be easily seen that $\hat\ba$ is an interior point of the uncertainty set ${\cal A}_1$.
Let
\begin{equation*}%\label{strict-feasible-soln-primal-SDP}
\bX(\lambda)=(1-\lambda)\left[\begin{array}{cc}\hat\ba\hat\ba^H& \hat    \ba\\ \hat\ba^H& 1\end{array}\right] + \lambda \left[\begin{array}{cc}\bI&\bzero\\ \bzero& 1\end{array}\right],\,\lambda\in(0,1) .
\end{equation*}
It can be observed that $\bX(\lambda)\succ\bzero$ for $0<\lambda<1$. It is also easy to check that $\tr(\bA_1\bX(\lambda))=-(1-\lambda)\epsilon+\lambda(\|\hat\ba\|^2-\epsilon+N)$, $\tr(\bA_2\bX(\lambda))=N$, and $\tr(\bA_3\bX(\lambda))=1$ for any $\lambda\in(0,1)$. Therefore, for sufficiently small $\lambda>0$, we have $\tr(\bA_1\bX(\lambda))<0$, and the matrix $\bX(\lambda)\succ\bzero$ is strictly feasible for SDP \eqref{min-constraint-homo-SDR}.

To construct a strictly feasible point for the dual SDP \eqref{min-constraint-homo-SDR-dual}, we note the constraint
\[
\left[\begin{array}{cc}\bw\bw^H-(y_1+y_2+y_3)\bI& y_1\hat\ba\\ y_1\hat\ba^H& -y_4-y_1(\|\hat\ba\|^2-\epsilon)\end{array}\right]\succeq\bzero.
\]
Let $y_2=1$, $y_3=-1$, $y_1$ be any negative number, and $y_4$ be such that
\[
-y_4>y_1(\|\hat\ba\|^2-\epsilon)+y_1^2\hat\ba^H(\bw\bw^H-y_1\bI)^{-1}\hat\ba.
\]
Then such a feasible point is strictly feasible for dual SDP \eqref{min-constraint-homo-SDR-dual}. Thus, we complete the proof.
\endproof

%\subsection{Proof of Lemma \ref{2nd-large-eigv-nonnegatiive}}\label{proof-2nd-large-eigv-nonnegatiive}
%\proof The second constraint of \eqref{rob-beamf-QMI} allows for the constraint
%\[
%\bw\bw^H-(y_1+y_2+y_3)\bI\succeq \bzero.
%\]
%We also know that $\lambda_i(\bA)\ge\lambda_i(\bB)$ for $\bA-\bB\succeq\bzero$, $i=1,\ldots,N$, where $\lambda_1(\bA)\ge\cdots \ge\lambda_N(\bA)$, and $\lambda_1(\bB)\ge\cdots\ge \lambda_N(\bB)$. Comparing the second large eigenvalues of $\bw\bw^H$ and $(y_1+y_2+y_3)\bI$, we then straightforwardly obtain \eqref{additional-linear-constr}.
%\endproof

\subsection{Proof of Proposition \ref{solution-property-1}}\label{proof-solution-property-1}
\proof
(1) Suppose that $y_1=0$. Then we have $y_2+y_3\le0$, and
\[
\left[\begin{array}{cc}\bW-(y_2+y_3)\bI& \bzero \\ \bzero & -y_4\end{array}\right]\succeq\bzero,
\]
which implies that $-y_4\ge0$. From the first constraint for the primal SDP, it follows that
\[
(N-\eta_1)y_2+(N+\eta_2)y_3-1=-y_4\ge0,
\]
which means further that
\[
0\ge\eta_2 y_3\ge 1+\eta_1 y_2 -N(y_2+y_3)\ge1.
\]
This is a contradiction, and thus, we have $y_1<0$.

(2) The third and the fourth constraints of problem \eqref{rob-beamf-QMI-1-LMI-restricted-dual} allow for the constraints
\begin{equation}\label{z0-upper-lower-bounds}
\frac{\tr\bZ-x}{N+\eta_2}\le z_0\le\frac{\tr\bZ-x}{N-\eta_1}.
\end{equation}
Therefore, on one hand, $\tr\bZ-x>0$ (otherwise $z_0\le0$, which clearly is impossible according to \eqref{strong-duality-two-LMIs}) while on the other hand, at least one of the two inequality in \eqref{z0-upper-lower-bounds} is strict. It follows from complementary conditions \eqref{QMI-complementary-conditions-3} and \eqref{QMI-complementary-conditions-4} that at least one of the two numbers $y_2$ and $y_3$ is zero. In other words, $y_2y_3=0$.

(3) It follows from the assumption that $y_2+y_3=-y_1>0$. Considering $y_2\ge0$ and $y_3\le0$, we have $y_2>0$. Therefore, it holds that $y_3=0$, since $y_2y_3=0$.
\endproof

\subsection{Proof of Proposition \ref{solution-property-2}}\label{proof-solution-property-2}
\proof
(1) According to \eqref{QMI-complementary-conditions-2} and the established fact that $y_1<0$ (the first claim in Proposition~\ref{solution-property-1}), we immediately have $\tr\bZ-2\Re(\hat\ba^H\bz_1)+z_0(\|\hat\ba\|^2-\epsilon )=x$, and then obtain $$z_0=\frac{2\Re(\hat\ba^H\bz_1)-(\tr\bZ-x)}{\|\hat\ba\|^2-\epsilon}.$$ In addition, it follows from \eqref{QMI-complementary-conditions-1}, \eqref{QMI-complementary-conditions-2}, \eqref{QMI-complementary-conditions-6}, \eqref{QMI-complementary-conditions-6-1} and \eqref{strong-duality-two-LMIs} that
\[
z_0=\frac{y_1x+(y_2+y_3)\tr\bZ}{1-y_4},
\]as long as $y_4\ne1$.

(2) If
\[
-y_4-y_1(\|\hat\ba\|^2-\epsilon)=0,
\]
then $y_1=0$ since $\bQ$ in \eqref{QMI-complementary-conditions-6-1} is positive semidefinite. This is a contradiction to the established fact that $y_1<0$ (the first claim in Proposition~\ref{solution-property-1}), and therefore, we have $$-y_4-y_1(\|\hat\ba\|^2-\epsilon)>0.$$

(3) From the first constraint in problem \eqref{rob-beamf-QMI-1-LMI-restricted} and the earlier established fact that $y_3=0$ (the third claim in Proposition~\ref{solution-property-1}), we have
\[
-y_4=y_2(N-\eta_1)-1.
\]
According to the previous claim, we obtain
\[
y_2(N-\eta_1)-y_1(\|\hat\ba\|^2-\epsilon)>1.
\]
It follows from the fact that $-y_1=y_2$ (also the third claim in Proposition~\ref{solution-property-1}) that
\[
y_2>\frac{1}{2N-\epsilon-\eta_1}.
\]
\endproof

\subsection{Proof of Theorem \ref{solution-property-3-necessary-condition-1}}\label{proof-solution-property-3-necessary-condition-1}
\proof
(1) It follows from \eqref{QMI-complementary-conditions-1} that $\bw^H(\hat\bR-\bZ)\bw=0$, which immediately implies that $(\hat\bR-\bZ)\bw=\bzero$ since $\hat\bR-\bZ\succeq\bzero$.

(2) Recalling \eqref{LMI-constraint-reformulate-Schur-complement}, we obtain
\begin{equation}\label{LMI-constraint-reformulate-Schur-complement-1}
-(y_1+y_2+y_3)\bI\succeq \bar\ba\bar\ba^H-\bw\bw^H.
\end{equation}
It can be verified that if $\bar\ba$ and $\bw$ are linearly independent, then $\bar\ba\bar\ba^H-\bw\bw^H$ is a rank-two indefinite matrix with eigenvalues
\begin{equation}\label{eigenvalues-rk2-indefinite}
\lambda=\frac{\|\bar\ba\|^2-\|\bw\|^2\pm\sqrt{(\|\bar\ba\|^2+\|\bw\|^2)^2-4|\bar\ba^H\bw|^2}}{2}.
\end{equation}
Therefore, due to \eqref{LMI-constraint-reformulate-Schur-complement-1}, we have
\[
-(y_1+y_2+y_3)\ge\frac{\|\bar\ba\|^2-\|\bw\|^2 + \sqrt{(\|\bar\ba\|^2+\|\bw\|^2)^2-4|\bar\ba^H\bw|^2}}{2},
\]
which is the positive eigenvalue of $\bar\ba\bar\ba^H-\bw\bw^H$, and thus, the proof is complete.
\endproof

\subsection{Proof of Theorem \ref{suffi-conditions-2}}\label{proof-suffi-conditions-2}
\proof
Let $(\tr(\hat\bR\bW),\tr(\bW),\tr(\bar\ba\bar\ba^H\bW))=(z_0,z_1,z_2)$, where in fact $z_0$ is the optimal value. Observe that
\[
\tr\left(\left(\hat\bR-\frac{z_0}{z_1}\bI\right)\bW\right)=0
\]
and
\[
\tr\left(\left(\bar\ba\bar\ba^H-\frac{z_2}{z_1}\bI\right)\bW\right)=0.
\]
It follows from Lemma~\ref{dcmp2} that there exists a rank-one matrix decomposition $\bW=\sum_{r=1}^R\bw_r\bw_r^H$ such that
\[
\tr\left(\left(\hat\bR-\frac{z_0}{z_1}\bI\right)\bw_r\bw_r^H\right)=0
\]
and
\[
\tr\left(\left(\bar\ba\bar\ba^H-\frac{z_2}{z_1}\bI\right)\bw_r\bw_r^H\right)=0,\,\forall r.
\]
Let
\begin{equation}\label{opt-soln-decom}
\bw=\sqrt{z_1}\frac{\bw_1}{\|\bw_1\|}.
\end{equation}
Then, it is not hard to verify  that
\begin{equation}\label{three-terms-matched}
(\bw^H\hat\bR\bw,\|\bw\|^2,|\bar\ba^H\bw|^2)=(z_0,z_1,z_2).
\end{equation}
Therefore, we claim that assumption \eqref{suffi-conditions-1-assumption-2} leads to
\begin{equation}\label{suffi-conditions-1-assumption-2-0}
y^2+y(\|\bar\ba\|^2-\|\bw\|^2)-(\|\bw\|^2\|\bar\ba\|^2-|\bar\ba^H\bw|^2)\ge0.
\end{equation}
Hence, it follows that
\begin{equation}\label{QMI-feasibility}
\bar\ba^H(-(y_1+y_2+y_3)\bI+\bw\bw^H)^{-1}\bar\ba\le1.
\end{equation}
In other words,
\begin{equation}\label{QMI-feasibility-1}
\left[\begin{array}{cc}\bw\bw^H-(y_1+y_2+y_3)\bI& y_1\hat\ba\\ y_1\hat\ba^H& -y_4-y_1(\|\hat\ba\|^2-\epsilon )\end{array}\right]\succeq\bzero.
\end{equation}
This fact together with \eqref{three-terms-matched} implies that $\bw\bw^H$ (with $\bw$ defined by \eqref{opt-soln-decom}) is optimal for SDP problem \eqref{rob-beamf-QMI-1-LMI-restricted}.

As for how to get \eqref{QMI-feasibility} from \eqref{suffi-conditions-1-assumption-2-0}, we calculate it as follows. Using the equality
\[
\begin{array}{l}
(-(y_1+y_2+y_3)\bI+\bw\bw^H)^{-1}\\
~~~~~~~~~~~~~~=\frac{-1}{y_1+y_2+y_3}\bI-\frac{\frac{-1}{y_1+y_2+y_3}\bw\bw^H}{-(y_1+y_2+y_3)+\bw^H\bw},
\end{array}
\]
it is straightforward to see that \eqref{QMI-feasibility} is equivalent to \eqref{suffi-conditions-1-assumption-2-0}.
\endproof

\subsection{Proof of Proposition \ref{2-solution-property-1}}\label{proof-2-solution-property-1}
\proof
(1) Suppose that $x=0$. Then $y_1+y_2<0$. From \eqref{QMI-2-complementary-conditions-1} and \eqref{QMI-2-complementary-conditions-6}, we then can see that
\[
(y_1+y_2)\tr \bZ=\tr(\bW\bZ)=\tr(\hat\bR\bW)>0,
\]
which is a contradiction. Here we use the fact that the optimal values $\tr(\hat\bR\bW)=z_1>0$. Therefore, we necessarily have $x<0$.

(2) Using the assumption that $x\lambda_{N-1}(\bar\bC)+y_1+y_2<0$, it follows from \eqref{QMI-2-complementary-conditions-5} that $z_2=0$, and it also follows from \eqref{QMI-2-complementary-conditions-2} and the above established fact that $x<0$ that $z_1=\frac{\tr(\bar\bC\bZ)}{\Delta_0}$.

(3) From \eqref{QMI-2-complementary-conditions-6}, we have
\begin{eqnarray*}
(y_1+y_2)\tr\bZ&=&\tr(\bZ\bW)-x\tr(\bar\bC\bZ)\\
&=&z_1-x\Delta_0 z_1\\
&=&z_1(1-x\Delta_0)\\
&>&0,
\end{eqnarray*}
where we apply $\tr(\bZ\bW)=\tr(\hat\bR\bW)=z_1$ and $\tr(\bar\bC\bZ)=z_1\Delta_0$. Hence,  $y_1+y_2>0$. %According to the assumption that $x\lambda_{N-1}(\bar\bC)+y_1+y_2<0$, we have $y_1+y_2<-x\lambda_{N-1}(\bar\bC)$.
\endproof

\subsection{Proof of Proposition \ref{2-solution-property-2}}\label{proof-2-solution-property-2}
\proof
(1) It follows from \eqref{QMI-2-complementary-conditions-1} and \eqref{QMI-2-complementary-conditions-6} that
\begin{eqnarray*}
0<\tr(\hat\bR\bW)=\tr(\bZ\bW)&=&\tr((x\bar\bC+(y_1+y_2)\bI)\bZ)\\
                             &=&x\tr((\bar\bC-\lambda_{N-1}(\bar\bC)\bI)\bZ).
\end{eqnarray*}
Since $x\le0$, hence we have $x<0$, and
\begin{equation}\label{upperbound-1}
\tr(\bar\bC\bZ)<\lambda_{N-1}(\bar\bC)\tr\bZ.
\end{equation}

(2) According to \eqref{QMI-2-complementary-conditions-2}, the optimal value is $$z_1=\frac{\tr(\bar\bC\bZ)+z_2\lambda_{N-1}(\bar\bC)}{\Delta_0}>0.$$

(3) %From the previous statement, we have
%\begin{eqnarray*}
%0<z_1&=&\frac{\tr(\bar\bC\bZ)+z_2\lambda_{N-1}(\bar\bC)}{\Delta_0}\\
%     &<&\frac{\lambda_{N-1}(\bar\bC)\tr\bZ+z_2\lambda_{N-1}(\bar\bC)}{\Delta_0}\\
%     &=&\frac{\lambda_{N-1}(\bar\bC)(\tr\bZ+z_2)}{\Delta_0}\\
%     &\le&\frac{\lambda_{N-1}(\bar\bC)z_1(N+\eta_2)}{\Delta_0},\\
%\end{eqnarray*}
%where the first inequality is due to \eqref{upperbound-1} and the second inequality is due to the dual feasibility (the third constraint of \eqref{rob-beamf-QMI-A2-LMI-res-dual}). Therefore, it follows that
%\[
%0<\Delta_0\le\lambda_{N-1}(\bar\bC)(N+\eta_2),
%\]
%which means that $\lambda_{N-1}(\bar\bC)>0$. Thereby, $y_1+y_2=-x\lambda_{N-1}(\bar\bC)>0$.
From \eqref{upperbound-1}, it follows that
\[
0\le\tr(\bar\bC\bZ)<\lambda_{N-1}(\bar\bC)\tr\bZ,
\]since both $\bar\bC$ and $\bZ$ are PSD. Observing $\lambda_{N-1}(\bar\bC)\ge0$, we therefore have $\lambda_{N-1}(\bar\bC)>0$, which implies that $y_1+y_2=-x\lambda_{N-1}(\bar\bC)>0$.

(4) By the assumption, we have $y_1+y_2=-x\lambda_{N-1}(\bar\bC)$. From the second constraint of \eqref{rob-beamf-QMI-A2-LMI-res}, it follows that
\[
\bW-x\bar\bC+x\lambda_{N-1}(\bar\bC)\bI\succeq\bzero,
\]
which implies that
\begin{equation}\label{eqns-in-proof-1}
\bW-x\bar\bC\succeq-x\lambda_{N-1}(\bar\bC)\bI.
\end{equation}
Therefore, we can write that
\[
\lambda_1(\bW)-x\lambda_{N}(\bar\bC)\ge\lambda_{N}(\bW-x\bar\bC)\ge-x\lambda_{N-1}(\bar\bC),
\]
where we use the property $\lambda_1(\bA)+\lambda_n(\bB)\ge\lambda_n(\bA+\bB)$  in the first inequality and apply the fact that $\lambda_n(\bA) \ge \lambda_n(\bB)$ for $\bA\succeq\bB$ to \eqref{eqns-in-proof-1} in the second inequality.
In other words,
\[
\lambda_1(\bW)\ge-x(\lambda_{N-1}(\bar\bC)-\lambda_{N}(\bar\bC)).
\]

(5) It follows that
\[
\lambda_1(\bw\bw^H)=\|\bw\|^2\ge-x(\lambda_{N-1}(\bar\bC)-\lambda_{N}(\bar\bC))>0.
\]
Hence, we have
\[
\frac{\bw^H\hat\bR\bw}{-x(\lambda_{N-1}(\bar\bC)-\lambda_{N}(\bar\bC))}\ge\frac{\bw^H\hat\bR\bw}{\|\bw\|^2}\ge\lambda_N(\hat\bR),
\]which means that
\[
\bw^H\hat\bR\bw\ge-x(\lambda_{N-1}(\bar\bC)-\lambda_{N}(\bar\bC))\lambda_N(\hat\bR).
\]
%It follows from \eqref{eqns-in-proof-1} that
%\begin{equation}\label{eqns-in-proof-1-rank-1}
%\bw\bw^H-x\bar\bC\succeq-x\lambda_{N-1}(\bar\bC)\bI,
%\end{equation}
%which implies that
%\begin{equation}\label{eqns-in-proof-rank-1-1}
%\hat\bR^{1/2}\bw\bw^H\hat\bR^{1/2}-x\hat\bR^{1/2}\bar\bC\hat\bR^{1/2}\succeq-x\lambda_{N-1}(\bar\bC)\hat\bR.
%\end{equation}
%Therefore, we have
%\begin{equation}\label{eqns-in-proof-rank-1-2}
%\lambda_N\left(\hat\bR^{1/2}\bw\bw^H\hat\bR^{1/2}-x\hat\bR^{1/2}\bar\bC\hat\bR^{1/2}\right)\ge-x\lambda_{N-1}(\bar\bC)\lambda_N(\hat\bR).
%\end{equation}
%Note that
%\[
%\lambda_1(\hat\bR^{1/2}\bw\bw^H\hat\bR^{1/2})=\bw^H\hat\bR\bw,\,\lambda_N(\hat\bR^{1/2}\bar\bC\hat\bR^{1/2})=\lambda_N(\hat\bR\bar\bC).
%\]
%Hence,
%\begin{eqnarray*}
%& &\bw^H\hat\bR\bw-x\lambda_N(\hat\bR\bar\bC)\\
%& &\ge\lambda_N\left(\hat\bR^{1/2}\bw\bw^H\hat\bR^{1/2}-x\hat\bR^{1/2}\bar\bC\hat\bR^{1/2}\right)\\
%& &\ge-x\lambda_{N-1}(\bar\bC)\lambda_N(\hat\bR),
%\end{eqnarray*}which means that
%\begin{eqnarray*}
%\bw^H\hat\bR\bw&\ge& x\lambda_N(\hat\bR\bar\bC)-x\lambda_{N-1}(\bar\bC)\lambda_N(\hat\bR)\\
%               &=&-x(\lambda_{N-1}(\bar\bC)\lambda_N(\hat\bR)-\lambda_N(\hat\bR\bar\bC)).
%\end{eqnarray*}
\endproof

\subsection{Proof of Theorem \ref{sufficient-conditions-3}}\label{proof-sufficient-conditions-3}
\proof
It can be observed that if $\bZ$ is a rank-one matrix, then it follows from \eqref{QMI-2-complementary-conditions-1} that $\bW$ must be of rank one (since $\hat\bR$ is positive definite and $\bW\ne\bzero$). Now, we suppose that the rank of $\bZ$ is greater than one. Let $(\tr(\hat\bR^{-1}\bZ),\tr(\bar\bC\bZ),\tr(\bZ))=(\delta_1,\delta_2,\delta_3)$. Then, we have
\[
\tr\left(\left(\hat\bR^{-1}-\frac{\delta_1}{\delta_3}\bI\right)\bZ\right)=0,\,
\tr\left(\left(\bar\bC-\frac{\delta_2}{\delta_3}\bI\right)\bZ\right)=0.
\]
From Lemma \ref{dcmp2}, it follows that there is a rank-one matrix decomposition $\bZ=\sum_{r=1}^R \bz_r\bz_r^H$ (where $R$ is the rank of $\bZ$) such that
\[
\tr\left(\left(\hat\bR^{-1}-\frac{\delta_1}{\delta_3}\bI\right)\bz_r\bz_r^H\right)=0,\,
\tr\left(\left(\bar\bC-\frac{\delta_2}{\delta_3}\bI\right)\bz_r\bz_r^H\right)=0,
\]
$r=1,\ldots,R$.

Define
\[
\bz=\sqrt{\delta_3}\frac{\bz_1}{\|\bz_1\|}.%=\sqrt{\sum_{r=1}^R\|\bz_r\|^2}\frac{\bz_1}{\|\bz_1\|}.
\]
It can be verified that
\begin{equation}\label{three-quadratic-forms-equal}
(\bz^H\hat\bR^{-1}\bz,\bz^H\bar\bC\bz,\bz^H\bz)=(\tr(\hat\bR^{-1}\bZ),\tr(\bar\bC\bZ),\tr(\bZ)).
\end{equation}
Since $\tr(\hat\bR^{-1}\bZ)\le1$, hence $\bz^H\hat\bR^{-1}\bz\le1$, which implies that
\[
\left[\begin{array}{cc}\hat\bR&\bz\\ \bz^H&1\end{array}\right]\succeq\bzero,
\]
which in turn means that $\hat\bR-\bz\bz^H\succeq\bzero$. Therefore, it follows from \eqref{three-quadratic-forms-equal} that $(\bz\bz^H,z_1,z_2)$ is an optimal solution for problem \eqref{rob-beamf-QMI-A2-LMI-res-dual}, and thus, the optimal solution $(\bW,x,y_1,y_2)$ for primal problem \eqref{rob-beamf-QMI-A2-LMI-res} fulfills the complementary conditions \eqref{QMI-2-complementary-conditions-1}-\eqref{QMI-2-complementary-conditions-6}. Again from \eqref{QMI-2-complementary-conditions-1}, we claim that the rank of $\bW$ is one.
\endproof

\end{document}